\documentclass[useAMS,usenatbib]{mn2e}
\usepackage{epsfig}
\usepackage[fleqn]{amsmath}
\title[Simulations of substructures in relativistic formations jets]{Simulations of substructures in relativistic jets}
\author[R. O. Garcia and S. R. Oliveira]{R. O. Garcia $^1$ \thanks{E-mail:
gr.gubim@gmail.com} and S. R.
Oliveira $^1$ \thanks{E-mail: samuel@ime.unicamp.br} \\ \footnotemark[1]
$^1$ Institute of Mathematics, Statistics and Scientific Computing, Campinas 13083-859, Brazil}
\newcommand{\sign}{\operatorname{sign}}

\begin{document}


\pagerange{\pageref{firstpage}--\pageref{lastpage}} \pubyear{2015}

\maketitle

\label{firstpage}

\begin{abstract} 
We present a set of simulations of relativistic jets from accretion disk initial setup with a new code in Fortran 90 to get numerical solutions of a system of General Relativistic Magnetohydrodynamics (GRMHD) partial differential equations in a fixed Black Hole (BH) spacetime which is able to show substructures formations inside the jet as well as a lobe formation on the disk. For this, a central scheme of finite volume method without dimensional split and no Riemann solvers (a Nessyahu-Tadmor method) was implemented. Thus, we were able to obtain stable numerical solutions with spurious oscillations under control and no excessive numerical dissipation. We setup a magnetized accretion disk uncharged plasma surrounding a central Schwarzschild BH immersed in a magnetosphere which evolve to the ejection of matter in the form of jet with its substructures over a distance of almost twenty times the BH radius. 
\end{abstract}

\begin{keywords}
methods: numerical, accretion discs, relativistic processes, MHD, galaxies: jets
\end{keywords}

\section{Introduction}

In the last decade or so, many astrophysical systems have been discovered with the property of generating collimated flow of plasma with speeds that can approach that of light.

Eruption in form of jets is one of the visible appearances of active compact objects that releases large amounts of energy and affects several astrophysical processes in the jets neighbourhood.

The physics and dynamics of the formation and ejection of collimated plasma, in short and long distances with respect to the radius of the central compact object, are still not fully understood, despite great theoretical advances, high performance computer simulations, high precision astronomical observations and enormous experimental plasma researches in laboratories \citep{thorne}, \citep{blandford}, \citep{chakrabarti}, \citep{beskinIOP}, \citep{beskinB}, \citep{mignone}, \citep{koide}, \citep{mckinney}, \citep{nishikawa}, \citep{shibata}, \citep{doeleman},\citep{massi},\citep{worral},\citep{marti},\citep{ciardi}.

It is fairly known that strong toroidal magnetic fields from the accretion disk have theirs lines changed sufficiently to collimate, guide and contain the ejected plasma without spreading during its temporal evolution. In the past decades researches on jet stability have shown the existence of several unstable modes as well stability criteria by analytical techniques in order to stabilize the jet propagation, but so far nothing can fully explain the length of several observable jets \citep{bai}, \citep{font_b},  \citep{beskinB}.

The substructures formed during propagation jets are also intriguing. Previous studies show that the possible types and forms of the jets and theirs substructures are affected by the accretion disk properties, the magnetic field and the magnetosphere where the system is immersed in \citep{mckinney}. However, the settings of the initial data for the analytical and computational analysis are still poisoning to remain difficult for the comparison between the theory and the astrophysical observations \cite{worral} or the experiments in plasma laboratories \citep{ciardi}. 

Astrophysical systems whose central compact object attracts matter around it, have been modelled by the equations of general relativity and magnetohydrodynamics (GRMHD).

Simulations that consider together the accretion process, the jet formation and its ejection in the same temporal evolution have been recently studied by \citep{koide}, \citep{nishikawa}, \citep{mckinney}, \citep{shibata} and \citep{blandford}.
However, the computational implementations of these phenomena through numerical methods are still challenging mainly because it involves many orders of magnitudes in space and time and also because the equations are non-linear.

The moments of transition after the accretion disk formation and the beginning of the ejection are the most difficult to gasp in the numerical simulations because the abrupt changes of physical quantities and the inadequate numerical treatment which add spurious oscillations or excessive dissipation that impairs progress of the simulation or degrade completely the numerical results obtained.

In this context, \citep{garcia} developed a new GRMHD code applied to the formation of relativistic jets, able to describe it from the accretion disk of matter up to his ejection altogether. For such, a central scheme of Finite Volume without the usual dimensional decomposition and no need of Riemann problem solvers, namely Nessyahu-Tadmor method, was implemented. With this code it was possible to obtain stable numerical solutions - without spurious oscillations or excessive dissipation - from the magnetized accretion disk process in rotation with respect to a central Schwarzschild  black hole (BH)  immersed in a magnetosphere, for the ejection of matter in the form of jet.

We present in this article the settings of the initial conditions capable of simulating the formation of relativistic jets from the accretion disk of matter up to its ejection, including the substructures formations inside jet, through the code developed in \citep{garcia} after the initial tests found in \citep{garciaol}. Furthermore, we obtained stable jets that reached 20 times the Schwarzschild radius and we found some rare simulations with the formation of lobes on the disk itself.

This paper is structured as follows: in section 2 we present the basic equations needed for the modelling the jets formation; in section 3 we show numerical methods that were applied on problem in study through the new code developed in Fortran 90 - the numerical solutions were obtained through the two-dimensional Nessyahu-Tadmor method and the Euler method of four stages to the source terms; in section 4 we have a simulation of relativistic jet formation and lobe formation; and in section 5 we wrote the conclusions of this work.

\section{Modeling of relativistic jets}\label{modeljet}

\subsection{Scenario in study}\label{scenery}

We consider the GRMHD equations with predetermined gravitation by a central compact object described simply by a fixed Schwarzschild BH. 

Initially we have a magnetized accretion disk with a strong magnetic field spinning surrounding of the BH and both immersed in a magnetosphere in free fall with respect to BH. The disk initial set up employs a Keplerian velocity profile.
From that, the basic dynamic scenery is the following: the disk's fluid is directed into the BH taking with it its magnetic field lines. Then, the material deposited in the vicinity of BH is squeezed and the differential disk rotation velocities provides a helical magnetic field along which the matter is aligned.

Depending on the initial values and profile of the: magnetic field, accretion disk density, density of the magnetosphere, disk rotation and opening of disk; a variety of outcomes are possible as, for example: a large out radial wind not collimated, jets that flowing outwards of the BH poles or lobes on the disk. 

In case of jets, we find and confirm several studies that the jets motion creates a longitudinal electrical current inducing a toroidal magnetic field able to provide a magnetic pressure that contributes to the jet collimation.

Consequently, the jet is formed on the poloidal region of BH, such that its collimation is maintained by the evolved magnetic field. Furthermore, our simulations show some substructures which are formed in the jet during its evolution.

For some initial conditions one can obtain collimated jets, whose extreme in contact with the magnetosphere, concentrates the matter in the form of a lobe in a well understood shock wave pattern.

In the following subsections we describe the equations of the mathematical model, the settings and initial conditions to for the full evolution from the accretion disk to the formation of relativistic jets and its substructures.

\subsection{Basic equations}\label{modeling}

The ideal GRMHD equations for modelling the scenery described in section \ref{scenery} are the following \citep{thorne}:
\begin{align}
\nabla_{\mu} (\rho U^{\mu})&=0 \, , \label{mn0001} \\
\nabla_{\mu} T_g ^{\mu\nu}&=0 \; , \\
\partial_{\mu} F_{\nu\lambda} + \partial_{\nu} F_{\lambda\mu} + \partial_{\lambda} F_{\mu\nu}&=0 \; , \\
\nabla_{\mu} F^{\mu\nu}&=0 \; , \label{mn0004}
\end{align}
where  $\nabla_{\mu}$ represent the covariant derivative, $\partial_{\mu}$ the usual partial derivative,  $U^{\mu}$ the fluid's four-velocity,  $\rho$ the fluid's rest mass density, $T_g ^{\mu\nu}$  the of the energy momentum tensor of both fluid and electromagnetic field; and $F^{\mu\nu}$ is the electromagnetic field tensor. The components of the energy-momentum tensor are given by
\[ 
T_g ^{\mu\nu} =pg^{\mu\nu}+(e + p)U^{\mu}U^{\nu}+F_{\sigma}^{\mu}F^{\nu\sigma}- \frac{1}{4}g^{\mu\nu}F^{\lambda\kappa}F_{\lambda\kappa} \, , 
\]
where $g^{\mu \nu}$ is associate to the space-time metric $g_{\mu \nu}$ as usual, namely $g^{\mu \nu} g_{\nu \lambda} = \delta^\mu _\lambda $, the $4\times 4 $ unity matrix.  
In the case of a diagonal metric satisfying $g_{\mu\nu}=0$ for $\mu\neq\nu$, we set $h_{0}=\sqrt{-g_{00}}$, $h_{1}=\sqrt{g_{11}}$, $h_{2}=\sqrt{g_{22}}$ and $h_{3}=\sqrt{g_{33}}$.
The scalars $\rho$, $p$ and $e$ are the proper mass density, proper pressure and proper total energy density
$e=\rho + \frac{p}{\Gamma -1}$, 
respectively. The constant $\Gamma$ is related to the fluid specific heat and, unless otherwise stated, we set the light speed $c=1$ for simplicity.

The equations (\ref{mn0001})-(\ref{mn0004}) are rewritten in a conservation law form. In doing so we can use finite volume methods for obtaining good numerical solutions. This formulation was initially developed by Koide, Kudoh and Shibata \citep{shibata} and some generalization of it are discussed in \citep{font_a}. 

In this formulation, the components of the vectors $\textbf{v}$ of velocity, $\textbf{B}$ of magnetic field and $\textbf{E}$ of electric field, in fiducial coordinates, are defined by 
\begin{align}
v_{i} = \frac{h_{i}}{\gamma}U^{i} \; , \textrm{fixed } i \;, \\
B_{i} = \epsilon_{ijk}\frac{h_{i}}{J}F^{jk} \; , \\
E_{i} = \frac{1}{h_{0}h_{i}}F^{0i} \; ,\textrm{fixed } i \;,
\end{align}
in which $\gamma$ is the Lorentz factor, $ \epsilon_{ijk}$ is the permutation symbol and $J=h_{1}h_{2}h_{3}$ is the Jacobian of a coordinate transformation.

The quantities that are appropriate for the conservation law like equations,  in fiducial coordinates, are given by the mass density and total energy density and linear momentum density:
\begin{align}
 D = \gamma \rho \; , \\
 \mathcal{E} = (e + p)\gamma^{2} - p - D +  \frac{1}{2} \left( B^{2}+ E^{2} \right) \; , \\
 \textbf{P} = \left[ (e+p)\gamma^{2}\textbf{v} + \textbf{E}\times\textbf{B} \right] \; ,
\end{align}
respectively; and the stress-tension tensor
\begin{equation}
 T = p\textbf{I}+\gamma^{2} \left(e+p\right)\textbf{v}\textbf{v} - \textbf{B}\textbf{B} 
 -\textbf{E}\textbf{E} + \frac{1}{2} \left( B^{2}+ E^{2} \right)\textbf{I},
\end{equation}
with the components being computed from
\[ T^{ij} = h_{i}h_{j}T_{g}^{ij} \textrm{ , fixed } i,j=1,2,3 \; . \]

So, the equations (\ref{mn0001})-(\ref{mn0004}) are rewritten in the following form:
\begin{equation}
 \frac{\partial\textbf{u}}{\partial t} = 
 -q \circ \left[ \sum_{i=1}^{3} \frac{\partial}{\partial x^{i}}\left( h_{0}b_{i}\circ w_{i} \right) \right] + h_{0}f \, ,
 \label{edpconser}
\end{equation}
in which
\[ 
\textbf{u}= \left( D , \textbf{P}, \mathcal{E}, \textbf{B}\right)^T; 
\]
the circle $\circ$ is defined by
\begin{displaymath}
 a \circ b = \left(\begin{array}{c}
              a_{1} \\
              \vdots \\
              a_{n}
             \end{array}\right) \circ 
             \left(\begin{array}{c}
              b_{1} \\
              \vdots \\
              b_{n}
             \end{array}\right) =
              \left(\begin{array}{c}
              a_{1}b_{1} \\
              \vdots \\
              a_{n}b_{n}
             \end{array}\right)
\end{displaymath}
with
\begin{displaymath}
          q = \left( \begin{array}{c}
                      1/J \\ 1/J \\ 1/J \\ 1/J \\ 1/J \\ h_{1}/J \\ h_{2}/J \\ h_{3}/J
                     \end{array}  \right) , \;
 (\;b_{1}\;|\;b_{2}\;|\;b_{3}\;) = \left( \begin{array}{ccc}
                      h_{2}h_{3} & h_{3}h_{1} & h_{1}h_{2} \\
                      h_{2}h_{3} & h_{3}h_{1} & h_{1}h_{2} \\
                      h_{2}h_{3} & h_{3}h_{1} & h_{1}h_{2} \\
                      h_{2}h_{3} & h_{3}h_{1} & h_{1}h_{2} \\
                      h_{2}h_{3} & h_{3}h_{1} & h_{1}h_{2} \\                     
                      0 & h_{3} & h_{2} \\
                      h_{3} & 0 & h_{1} \\
                      h_{2} & h_{1} & 0 \\  
                     \end{array}  \right)        
\end{displaymath}
and
 \begin{displaymath}
 (w_{1}|w_{2}|w_{3}) = \left( \begin{array}{ccc}
                      Dv_{1} & Dv_{2} & Dv_{3} \\
                      T^{11} & T^{12} & T^{13} \\
                      T^{21} & T^{22} & T^{23} \\
                      T^{31} & T^{32} & T^{33} \\
                      (P_{1} -Dv_{1}) & (P_{2}-Dv_{2}) & (P_{3}-Dv_{3}) \\                     
                      0 & E_{3} & -E_{2} \\
                      -E_{3} & 0 & E_{1} \\
                      E_{2} & -E_{1} & 0 \\  
                     \end{array}  \right)         
\end{displaymath}
The source term of the equation (\ref{edpconser}) is given by
 \begin{displaymath}
 f = \left( \begin{array}{c}
                      0 \\
                      (\mathcal{E} + D)H_{01} + H_{12}T^{21} + H_{13}T^{31} - H_{21}T^{22} - H_{31}T^{33} \\
                      (\mathcal{E} + D)H_{02} + H_{23}T^{32} + H_{21}T^{12} - H_{32}T^{33} - H_{12}T^{11} \\
                      (\mathcal{E} + D)H_{03} + H_{31}T^{13} + H_{32}T^{23} - H_{13}T^{11} - H_{23}T^{22}\\
                      (H_{01}P_{1} + H_{02}P_{2} + H_{03}P_{3}) \\                     
                      0  \\
                      0  \\
                      0 \\  
                     \end{array}  \right)       
\end{displaymath}
with the auxiliar tensor $H_{\mu\nu}$ whose components are defined by  
\begin{equation}
 H_{\mu\nu} = -\frac{1}{h_{\mu}h_{\nu}}\left( \frac{\partial h_{\mu}}{\partial x^{\nu}} \right) \, ,
\end{equation}
where $\mu,\nu=0,1,2,3$.

The fluid is considered free of electromagnetic forces, no electrical resistivity, in the approximation known as the \textit{frozen-in} condition, given by
\begin{equation}
 \textbf{E} = -\textbf{v}\times \textbf{B} \, .
\end{equation}

\subsection{Schwarzschild metric}\label{schwarzschild}

The problem proposed of relativistic jets is modelled by the free-fall, orbit and flow of a magnetized perfect fluid in the static spherical symmetric Schwarzschild spacetime such that the BH is responsible for all gravitation involved on the problem. Therefore, in this model, the fluid has no self-gravitation. The spacetime metric is given by
\begin{equation}
ds^{2}\;=\; -\alpha^{2}dt^{2}+\frac{1}{\alpha^{2}}dr^{2}+r^{2}d\theta^{2}+r^{2}\sin^{2}\theta d\phi^{2}
\end{equation}
in which $\alpha$ is the lapse function
\[ \alpha = \sqrt{1-\frac{r_{s}}{r}} , \]
$r_{s}=2M$ is the Schwarzschild radius used as distance unit and $M$ is the BH mass. The gravitational constant is set to ${G=1}$. The coordinates $ (t, r, \theta, \phi) $ are respectively time, the radial coordinate, the polar and the azimuthal coordinate and thus
\[ h_{0} = \alpha, \;\;\; h_{1}=\frac{1}{\alpha}, \;\;\; h_{2}=r\, ,\;\;\;\;h_{3}=r\sin\theta\, . \]

\subsection{Accretion disk}\label{disk}

Simulations with stable jets formations are obtained through an initial accretion disk which is geometrically thin \citep{komissarov,mckinney} in Keplerian rotation velocity profile in azimuthal direction  \citep{beskinB}, that is,  
\begin{equation}
 v_{\phi}=v_{K} \equiv 1/{\sqrt{2\left( \frac{r}{r_{s}} - 1 \right)}}\, .
\end{equation}
The accretion disk is a magnetized fluid, which is initially localized on
\begin{equation}
 |\tan\bar{\theta}| < \delta \;\;\;\;\textrm{and}\;\;\;\; r > r_{D}=d_{in}r_{s}
\end{equation}
where $d_{in}>1$ is the position of the internal edge whose rotation velocity is

$$ v_{K}=1/{\sqrt{2(d_{in}-1)}}\; $$
and $ \bar{\theta}=\dfrac{\pi}{2} - \theta \in [-\pi/2,\pi/2]$ is the angle with respect to the equator.
A value studied for the opening of the disk is $\delta = 1/8 $.

The magnetosphere that surrounds the system compound of the accretion disk and the BH is less dense than the disk and, initially, it free-falls toward the BH containing, so only the velocity's radial component is non-zero. Thus, the initial conditions on the density and velocity of matter around the BH are given by
\begin{flushleft}
 \textbf{Density:}
\end{flushleft}
\begin{equation}
 \rho = \rho_{mag} + \rho_{disk}
\end{equation}
where $\rho_{mag}$ and and $\rho_{disk}$  are the fluid's density of the magnetosphere and of disk, respectively and
\begin{equation}
  \rho_{disk}=\left\lbrace \begin{array}{ll}
                              \kappa_{\rho}\rho_{mag}, & \textrm{if} \;r>r_{D}\;\;\textrm{and}\;\;|\tan\bar{\theta}|<\delta  \\
                              0, & \textrm{if} \;r\leq r_{D}\;\;\textrm{or} \;\;|\tan\bar{\theta}|\geq\delta
                             \end{array}\right. \;\, ,
\end{equation}
in which $\kappa_{\rho}$ is ratio between disk density and magnetosphere density.
\begin{flushleft}
 \textbf{Velocity:}
\end{flushleft}
\begin{equation}
 (v_{r},v_{\theta},v_{\phi})=\left\lbrace \begin{array}{l}
                              (0,0,v_{K}), \;\;\; \textrm{if}\;r>r_{D}\; \textrm{and} \;|\tan\bar{\theta}|<\delta  \\
                              (-v_{mag},0,0), \; \textrm{if} \;r\leq r_{D}\;\textrm{or} \;|\tan\bar{\theta}|\geq\delta
                             \end{array}\right. ,
\end{equation}
where $v_{K}$ is the disk's Keplerian velocity, $v_{mag}$ is the in falling velocity of the magnetosphere's fluid.
\begin{flushleft}
 \textbf{Magnetic field:}
\end{flushleft}
\begin{equation}
 \left\lbrace \begin{array}{l}
                              B_{r} = B_{D}\cos\theta \\
                              B_{\theta} = -\alpha B_{D}\sin\theta \\
                              B_{\phi}=0
                             \end{array}\right.\;\, ,
\end{equation}
in which $B_{D}=\kappa_{B}\sqrt{\rho_{D}}$, $\rho_{D}$ is the density at the inner edge of the disk, $\kappa_{B}$ is a constant of proportionality between the $\rho_{D}$ and the magnitude of magnetic field \citep{wald}.

The magnetosphere region has only the radial component of velocity $v_{r}=-v_{mag}$. One way to determine $v_{mag}$ is through the following equation \citep{shibata},

\begin{equation}
 \alpha = \frac{H(\gamma^{-2}+\Gamma - 2)}{(\Gamma-1)\gamma},
 \label{mod010}
\end{equation}
where H is a constant related to the enthalpy of the fluid, $\Gamma$ is a constant related to the specific heat. Equation (\ref{mod010}) ensures that there is a sonic point between the values ​​of $\gamma$ satisfying the equality. This point separates regions where the fluid has transonic and subsonic speeds -- this property makes a more realistic accretion disk model  \citep{beskinB}. Solving equation (\ref{mod010}) 
we find values of $\gamma$ to each $\alpha$ and so $v_{mag}$ is determined by the Lorentz factor, that is, $ \gamma = 1/\sqrt{1-v_{mag} ^2} $.

Let the plasma be a polytropic gas, that is, $p=\rho^{\Gamma}$ and let the ratio between density and pressure be given by \cite{beskinB,shibata},
\begin{equation}
 a \equiv \frac{p}{\rho}=\frac{\Gamma-1}{\Gamma}\left( \frac{H}{\alpha\gamma} -1 \right)\, .
 \label{denspress}
\end{equation}
Therefore the expressions of density $\rho_{mag}$ and pressure are respectively,
\begin{equation}
 \rho_{mag} = a^{1/(\Gamma -1)}\;\;\;\;\; \textrm{and} \;\;\;\;\; p=a^{1\;+\;1/(\Gamma -1)}\,
\end{equation}
in which we used equation (\ref{denspress}).

We have almost all the information for initial setting. It remains to set the \textit{equation of state}. 

\subsection{Equation of State}\label{secEOS}

The equation of state is a system of algebraic equations:
\begin{equation}
  \begin{array}{l}
    x(x+1)\left[ \Gamma ax^{2} +(2\Gamma a-b)x + \Gamma a -b +d \frac{\Gamma}{2}y^{2} \right]^{2}= \\
    = \left(\Gamma x^{2}+2\Gamma x +1 \right)^{2}\left[ P^{2}(x+1)^{2}+ 2\sigma y +2\sigma xy + B^{2}y^{2} \right], \\
    \     \ \\
    \left[ \Gamma(a-B^{2})x^{2} + (2\Gamma a-2\Gamma B^{2}-b)x + \Gamma a -b +d - B^{2}\right]y + \\
    +\left[\frac{\Gamma}{2}y \right]y= \sigma(x+1)(\Gamma x^{2}+2\Gamma x +1)  ,
    \end{array}
\label{EOS}
\end{equation}
where $ x=\gamma-1$, $y=\gamma(\textbf{v}\cdot\textbf{B})$, $a=D+\mathcal{E}$, $b=(\Gamma-1)D$, $d=(1-\frac{\Gamma}{2})B^{2}$, $\sigma=\textbf{B}\cdot\textbf{P}$ in which $B$ and $P$ are the magnitude of magnetic field and momentum, respectively. Details on the equation of state (\ref{EOS}) are in \citep{mignone} and initial studies in \citep{schneider}. The values $ x $ and $ y $ are the unknowns for the equations (\ref{EOS}) which are solved by a Newton-Raphson method \citep{powell,press}. At each time step, the GRMHD equations require the solution of this system --  the the initial guess for the Newton-Raphson method is chosen as the solutions $ x $ and $ y $ of the previous time step. 

\subsection{Implemented Equations}\label{impeq}

The implemented equations in the new developed code are exposed in this section. From the equations of section \ref{modeljet} together with some numerical analysis hypotheses we set the numerical algorithms as discussed in \cite{garcia}.

In this work we assume axial symmetry so the spatial variables are: radial $r$ and polar $\theta$. Setting $x_{1}=r$, $x_{2}=\theta$ we get the following reduced equations:

\begin{flushleft}
 \textit{Mass Equation:}
\end{flushleft}
\begin{equation}
 \frac{\partial D}{\partial t} = -\frac{1}{J}\left\lbrace \frac{\partial}{\partial x^{1}}\left( h_{0}h_{2}h_{3}Dv_{1} \right)
    + \frac{\partial}{\partial x^{2}}\left( h_{0}h_{3}h_{1}Dv_{2} \right) \right\rbrace
\label{edp01}
\end{equation}

\begin{flushleft}
 \textit{Motion Equations:}
\end{flushleft}
\begin{equation}
 \frac{\partial P_{1}}{\partial t} = -\frac{1}{J}\left\lbrace \frac{\partial}{\partial x^{1}}\left( h_{0}h_{2}h_{3}T^{11} \right)
    + \frac{\partial}{\partial x^{2}}\left( h_{0}h_{3}h_{1}T^{12} \right) \right\rbrace + S_{2}
    \label{edp02}
\end{equation}
\begin{equation}
 \frac{\partial P_{2}}{\partial t} = -\frac{1}{J}\left\lbrace \frac{\partial}{\partial x^{1}}\left( h_{0}h_{2}h_{3}T^{21} \right)
    + \frac{\partial}{\partial x^{2}}\left( h_{0}h_{3}h_{1}T^{22} \right) \right\rbrace + S_{3}
\label{edp03}    
\end{equation}
\begin{equation}
 \frac{\partial P_{3}}{\partial t} = -\frac{1}{J}\left\lbrace \frac{\partial}{\partial x^{1}}\left( h_{0}h_{2}h_{3}T^{31} \right)
    + \frac{\partial}{\partial x^{2}}\left( h_{0}h_{3}h_{1}T^{32} \right) \right\rbrace + S_{4}
    \label{edp04}
\end{equation}

\begin{flushleft}
\textit{Energy Equation:}
\end{flushleft}

\begin{equation}
\begin{array}{l}
\displaystyle \frac{\partial \mathcal{E}}{\partial t} =  -\frac{1}{J}\left\lbrace\frac{\partial}{\partial x^{1}}\left[ h_{0}h_{2}h_{3}\left(P_{1}-Dv_{1}\right) \right] \right\rbrace+ \\
                \\                
\displaystyle -\frac{1}{J}\left\lbrace\frac{\partial}{\partial x^{2}}\left[ h_{0}h_{3}h_{1}\left(P_{2}-Dv_{2}\right)\right]  \right\rbrace + S_{5}
\end{array}                           
\label{edp05}
\end{equation}

\begin{flushleft}
 \textit{Equations of Magnetic Field:}
\end{flushleft}

\begin{equation}
 \frac{\partial B_{1}}{\partial t} = -\frac{h_{1}}{J}\left\lbrace \frac{\partial}{\partial x^{2}}\left( h_{0}h_{3}E_{3}\right) \right\rbrace
 \label{edp06}
\end{equation}

\begin{equation}
 \frac{\partial B_{2}}{\partial t} = -\frac{h_{2}}{J}\left\lbrace -\frac{\partial}{\partial x^{1}}\left( h_{0}h_{3}E_{3}\right) \right\rbrace
 \label{edp07}
\end{equation}

\begin{equation}
 \frac{\partial B_{3}}{\partial t} = -\frac{h_{3}}{J}\left\lbrace \frac{\partial}{\partial x^{1}}\left( h_{0}h_{2}E_{2}\right)
   - \frac{\partial}{\partial x^{2}}\left( h_{0}h_{1}E_{1} \right) \right\rbrace
   \label{edp08}
\end{equation}

\begin{flushleft}
\textit{Source Terms:}
\end{flushleft}

\begin{equation}
\begin{array}{rl}
 S_{2} = & h_{0}\left\lbrace \left( \mathcal{E} + D \right)H_{01} + H_{12}T^{21} + H_{13}T^{31}\right\rbrace + \\
         & - h_{0}\left\lbrace H_{21}T^{22} + H_{31}T^{33} \right\rbrace
\end{array}
 \label{tf01}
\end{equation}

\begin{equation}
\begin{array}{rl}
 S_{3} = & h_{0}\left\lbrace \left( \mathcal{E} + D \right)H_{02} + H_{23}T^{32} + H_{21}T^{12} \right\rbrace + \\
         & -h_{0}\left\lbrace H_{32}T^{33} + H_{12}T^{11} \right\rbrace
\end{array}
\end{equation}

\begin{equation}
\begin{array}{rl}
 S_{4} = & h_{0}\left\lbrace \left( \mathcal{E} + D \right)H_{03} + H_{31}T^{13} + H_{32}T^{23} \right\rbrace + \\
         & - h_{0}\left\lbrace H_{13}T^{11} + H_{23}T^{22} \right\rbrace
\end{array}
\end{equation}

\begin{equation}
 S_{5} = h_{0}\left\lbrace H_{01}P_{1} + H_{02}P_{2} + H_{03}P_{3} \right\rbrace
 \label{tf04}
\end{equation}

\subsection{Initial and boundary conditions}

The initial conditions must model the accretion disk along with the magnetosphere and the central BH. Thus, we implemented a modular structure in Fortran 90 specifically to deal with such a stage, which is the construction of the equations of sections \ref{modeling}, \ref{schwarzschild} and \ref{disk}.

The boundary conditions for the magnetosphere region between the inner and outer radii we used inward radiation conditions.

In outer edge, the disk is fed through matter, thus in this region we applied Rodin condition. To the boundaries $\theta=0$ e $\theta=\frac{\pi}{2}$, we used mirror symmetry.


\section{Choice of numerical methods}

Currently, the main codes used for the GRMHD equations have an emphasis on Godunov methods that need of analytical or approximate Riemann Solvers, with dimensional splitting and slope limiter, even though the equations system have degeneracy of eigenvalues obtained from the Jacobian matrix \citep{anninos}, \citep{coconut}, \citep{zanna}, \citep{font_b}, \citep{gammie}, \citep{giacomazzo2006}, \citep{mosta}, \citep{tchekhovskoy}. This limits the applicability of such methods. Another restrictive issue associated with these methods is the use of dimensional splitting, which aren't recommended for problems whose eigenvalues have large differences in orders of magnitude. These differences occur exactly at the transition that defines the formation of jets.

In this work, we used a central scheme without dimensional splitting and no need of Riemann Solvers, namely, two-dimensional Nessyahu-Tadmor method to obtain numerical solutions to the equations (\ref{edp01})-(\ref{edp08}). 

\subsection{Decomposition of source term}\label{termofonte}

The implemented equations (section \ref{impeq}) are the following form,
\[ u_{t}+f(u)_{x}+g(u)_{y}=s(u)\, . \]
Before applying the Nessyahu-Tadmor method, we separated the Partial Differential Equation (PDE) with source term at two systems of equations: an homogeneous PDE system without the source terms and an Ordinary Differential Equation (ODE) system with the source terms. Thus, between the time $t_{n}$ and $t_{n +1}$ we have the following problems:
\begin{equation}
 \left\lbrace \begin{array}{l}
               u_{t}+f(u)_{x}+g(u)_{y}=0 \\
               u(x,y,t_{n}) = u^{n}
              \end{array} \right. \Rightarrow \overline{u}^{n+1}
              \label{EDP05}
\end{equation}
and
\begin{equation}
 \left\lbrace \begin{array}{l}
               \frac{du}{dt}=s(u) \\
               u(x,y,t_{n}) = \overline{u}^{n+1}
              \end{array} \right. \Rightarrow u^{n+1} \;\, ,
              \label{EDO01}
\end{equation}
where $\overline{u}^{n+1}$ is a numerical solution for homogeneous PDE and equation (\ref{EDO01}) updates $\overline{u}^{n+1}$ to obtain a solution $u^{n+1}$ for the inhomogeneous ODE .

Therefore, to perform a time step from $ t_ {n} $ to $ t_ {n +1} $, we obtain an approximate solution to (\ref{EDP05}) and use this solution as the initial condition of the ODE.

The equation (\ref{EDP05}) is solved by Nessyahu-Tadmor method \citep{balbas} and the solution is updated into the equation (\ref{EDO01}) which is solved by a fourth-order Runge-Kutta method \citep{hunds}.

\subsection{Two-dimensional Nessyahu-Tadmor method} \label{NTbidimensional}

The Nessyahu-Tadmor method is a natural extension of the Lax-Friedrichs scheme and so it keeps the Lax-Friedrichs robustness providing stable solutions with no spurious oscillations nor excessive numerical dissipation \citep{nessyahu}.

In this subsection, we present the Nessyahu-Tadmor method in the bidimensional cartesian space domain.

Let $\Omega = \Omega_{x}\times\Omega_{y}$ be a regular partition of the space domain such that,
\[\Omega_{x}:x_{a}=x_{1}<\cdots<x_{i}<\cdots<x_{N+1}=x_{b}\]
and
\[\Omega_{y}:y_{a}=y_{1}<\cdots<y_{j}<\cdots<y_{M+1}=y_{b} ,\] with $N$ and $M$ subintervals in the $x$ and $y$ directions, respectively, in which, 
\[\Delta x=\frac{x_{b}-x_{a}}{N}\;\;\textrm{and}\;\;\Delta y=\frac{y_{b}-y_{a}}{M},\]
\[ x_i = x_a + (i-1) \Delta x \;\;\textrm{and}\;\;\ y_j = y_a + (j-1) \Delta y \, , \]
 $\forall i\in [1,N+1] $ and $\forall j\in [1,M+1] $. The $i,j$-th \textit{average cell} for the $i,j$-th \textit{finite volume} is defined by
$$ \Omega_{i,j}=(x_{i-1/2},x_{i+1/2})\times (y_{j-1/2},y_{j+1/2})$$ 
in which $ x_{i+½}=(x_{i+1}+x_{i})/2$ and $ y_{i+½}=(y_{i+1}+y_{i})/2$.

Let $u=u(x,y,t)$ be a function that represents a physical quantity defined in the spatial domain $\Omega$. If $u$ is conserved in $\Omega$, then the conservation law is satisfied on each finite volume $\Omega_{i,j}$, that is,
\begin{equation}
\begin{array}{l}
 \displaystyle \frac{d}{dt}\int_{\Omega_{i,j}}u(x,y,t)dx = \displaystyle f(u(x_{i+1/2},y_{j},t))  +  \\
                       \\
                                               \displaystyle - f(u(x_{i-1/2},y_{j},t)) + g(u(x_{i},y_{j+1/2},t)) + \\
                       \\                        
                                              \displaystyle   - g(u(x_{i},y_{j-1/2},t)).
\end{array}
  \label{ntb01}
\end{equation} 
When it define a regular partition for the independent variable $ t $ where $\Delta t =t_{n +1}-t_{n}$, an explicit algorithm for temporal evolution of a single step is obtained by integrating of equation (\ref{ntb01}) between $t_{n}$ and $t_{n+1}$.  Thus, we obtain the following equation,



\begin{equation}
\begin{array}{l}
\displaystyle \displaystyle \frac{1}{\Delta x \Delta y} \int_{\Omega_{i,j}}u(x,y,t_{n+1})dx =\\
      \\
 = \displaystyle  \displaystyle \frac{1}{\Delta x \Delta y}\int_{\Omega_{i,j}}u(x,y,t_{n})dx + \\
    \\
 +\displaystyle\frac{1}{\Delta x \Delta y}\int_{t_{n}}^{t_{n+1}}\int_{y_{j-1/2}}^{y_{j+1/2}}f(u(x_{i-1/2},y,t))dydt \; + \\
      \\
 -\displaystyle\frac{1}{\Delta x \Delta y}\int_{t_{n}}^{t_{n+1}} \int_{y_{j-1/2}}^{y_{j+1/2}}f(u(x_{i+1/2},y,t)dydt\; +  \\
      \\
+ \displaystyle \frac{1}{\Delta x \Delta y}  \int_{t_{n}}^{t_{n+1}} \int_{x_{i-1/2}}^{x_{i+1/2}} g(u(x,y_{j-1/2},t))dxdt + \\
      \\
- \displaystyle \frac{1}{\Delta x \Delta y}  \int_{t_{n}}^{t_{n+1}} \int_{x_{i-1/2}}^{x_{i+1/2}} g(u(x,y_{j+1/2},t))  dxdt. 
\end{array}
\label{ntb02}
\end{equation}
Equation (\ref{ntb02}) provides a procedure to find the values ​​of $u$ at time $t_{n+1}$, under points $(x_{i},y_{j})$. This integral equation is exact (no approximations) and is valid for any subintervals of $\Omega$.

As the aim is to get a Centred Finite Volume method free of Riemann Solvers, we consider the conservation law (\ref{ntb02}) on a staggered mesh, that is,
\[ I_{i,j}=(x_{i},x_{x_{i+1}})\times(y_{j},y_{j+1})\, , \]
thus, the equation (\ref{ntb02}) is rewritten as follows
\begin{equation}
\begin{array}{l}
\displaystyle \displaystyle \frac{1}{\Delta x \Delta y} \int_{I_{i,j}}u(x,y,t_{n+1})dx= \displaystyle  \displaystyle \frac{1}{\Delta x \Delta y}\int_{I_{i,j}}u(x,y,t_{n})dx + \\
      \\
 +\displaystyle\frac{1}{\Delta x \Delta y}\int_{t_{n}}^{t_{n+1}}\int_{y_{j}}^{y_{j+1}}\left[ f(u(x_{i},y,t))-f(u(x_{i+1},y,t) \right]  dydt\; +  \\
      \\
+ \displaystyle \frac{1}{\Delta x \Delta y}  \int_{t_{n}}^{t_{n+1}} \int_{x_{i}}^{x_{i+1}}\left[ g(u(x,y_{j},t)) - g(u(x,y_{j+1},t)) \right]  dxdt .
\end{array}
\label{ntb03}
\end{equation}
Now consider $w=w(x,y,t)$ a piecewise bilinear polynomial function in which
 \begin{equation}
  w(x,y,t_{n})=\sum_{i,j}\omega_{i,j}(x,y)p_{i,j}(x,y)\, ,
  \label{ntb04}
 \end{equation}
with $\overline{p}_{i,j}(x_{i},y_{j})=\overline{w}_{i,j}^{n}$ and $\omega_{i,j}=1$.
 
 Thus, the average value of $w(x,y,t_{n})$, defined on $I_{i,j}$ is given by
 \begin{equation}
 \begin{array}{l}
 \displaystyle \overline{w}_{i+1/2,j+1/2}^{n}= \displaystyle \frac{1}{\Delta x\Delta y}\int_{x_{i}}^{x_{i+1/2}} \int_{y_{j}}^{y_{j+1/2}} p_{i,j}(x,y)dydx +  \\
   \\
  \displaystyle +\frac{1}{\Delta x\Delta y}\int_{x_{i+1/2}}^{x_{i+1}}\int_{y_{j}}^{y_{j+1/2}}p_{i+1,j}(x,y)dydx + \\
                                  \\
                                \displaystyle +\frac{1}{\Delta x\Delta y} \int_{x_{i}}^{x_{i+1/2}}\int_{y_{j+1/2}}^{y_{j+1}}p_{i,j+1}(x,y)dydx + \\
                                  \\
                                
                                \displaystyle +\frac{1}{\Delta x\Delta y} \int_{x_{i+1/2}}^{x_{i+1}}\int_{y_{j+1/2}}^{y_{j+1}}p_{i+1,j+1}(x,y)dydx .
  \label{ntb05}
  \end{array}
 \end{equation}
If $w$ is an approximation to $u$, then from equations (\ref{ntb05}) and (\ref{ntb03}) we obtain
 \begin{equation}
\begin{array}{l}
\displaystyle \overline{w}_{i+1/2,j+1/2}^{n+1}=\overline{w}_{i+1/2,j+1/2}^{n} + \\
      \\
+\displaystyle\frac{1}{\Delta x \Delta y}\int_{t_{n}}^{t_{n+1}}\int_{y_{j}}^{y_{j+1}} f(w(x_{i},y,t))dydt + \\
 \\
-\displaystyle\frac{1}{\Delta x \Delta y}\int_{t_{n}}^{t_{n+1}}\int_{y_{j}}^{y_{j+1}}f(w(x_{i+1},y,t)  dydt\; +  \\
      \\
+ \displaystyle \frac{1}{\Delta x \Delta y}  \int_{t_{n}}^{t_{n+1}} \int_{x_{i}}^{x_{i+1}}g(w(x,y_{j},t)) dxdt + \\
     \\
- \displaystyle \frac{1}{\Delta x \Delta y}  \int_{t_{n}}^{t_{n+1}} \int_{x_{i}}^{x_{i+1}} g(w(x,y_{j+1},t)) dxdt \;\, .
\end{array}
\label{ntb06}
\end{equation}

Therefore, equation (\ref{ntb06}) provides an exact expression to the evolution of $w$ based on integration of the conservation law over the staggered cell $I_{i,j}$. However, we must obtain an expression for terms like $\overline{w}_{i+1/2,j+1/2}^{n}$ and fluxes' integrations. 

As $w$ is a piecewise bilinear polynomial, then
\begin{equation}
 p_{i,j}(x,y)=\overline{w}_{i,j}^{n} + \frac{D_{x}w_{i,j}}{\Delta x}\left(x - x_{i} \right) + \frac{D_{y}w_{i,j}}{\Delta y}\left(y - y_{j} \right)\, ,
 \label{ntb07}
\end{equation}
where 
$$D_{x}w_{i,j} = MM\left\lbrace \Delta_{x} w^{n}_{i+1/2,j}\, ,\;\Delta_{x} w^{n}_{i-1/2,j}\right\rbrace,$$ 
with 
$$\Delta_{x} w^{n}_{i+1/2,j} = w^{n}_{i+1,j} - w^{n}_{i,j} \, .$$
Similarly,
$$D_{y}w_{i,j} = MM\left\lbrace\Delta_{y} w^{n}_{i,j+1/2}\, ,\;\Delta_{y} w^{n}_{i,j-1/2}\right\rbrace$$ 
with 
$$\Delta_{y} w^{n}_{i,j+1/2} = w^{n}_{i,j+1} - w^{n}_{i,j}\,.$$

The symbol $MM\left\lbrace\cdot,\cdot\right\rbrace$ stands for \textit{minmod function} defined by
\begin{equation}
 MM\{a,b\} = \frac{1}{2}\left[\sign{a}+\sign{b} \right] \min \left\lbrace |a|,|b| \right\rbrace,
 \label{minmod}
\end{equation}
with $\sign{a}$ being the signal of number $a$.

Replacing  $p_{i,j}$ on equation (\ref{ntb05}), we obtain an expression for $\overline{w}_{i+1/2,j+1/2}^{n}$. Hence,
 \begin{equation}
 \begin{array}{l}
 \displaystyle \overline{w}_{i+1/2,j+1/2}^{n}= \displaystyle \frac{1}{4}\left(\overline{w}_{i,j}^{n}+\overline{w}_{i+1,j}^{n}+ \overline{w}_{i,j+1}^{n} +\overline{w}_{i+1,j+1}^{n}\right) + \\
                                                 \\
  \displaystyle + \frac{1}{16}\left[(D_{x}w_{i,j} - D_{x}w_{i+1,j}) + (D_{x}w_{i,j+1} - D_{x}w_{i+1,j+1}) \right] +  \\
                                                 \\
  \displaystyle + \frac{1}{16}\left[(D_{y}w_{i,j} - D_{y}w_{i,j+1}) +  (D_{y}w_{i+1,j} - D_{y}w_{i+1,j+1})\right]\;\, .
 \end{array}
 \label{ntb08}
 \end{equation}
 
Now that we know an expression for $\overline{w}_{i+1/2,j+1/2}^{n}$, it remains to find the integrations of the fluxes of the equation (\ref{ntb06}). For this, we used the middle point rule for the time integration and the trapezoidal rule for space integration \citep{powell}. Thus,
\begin{equation}
\begin{array}{l}
 \displaystyle \int_{t_{n}}^{t_{n+1}}\int_{y_{j}}^{y_{j+1}} f(w(x_{i},y,t))  dydt \cong \\
  \\
 \displaystyle \cong \frac{\Delta y \Delta t}{2}\left[ f\left( w^{n+1/2}_{i,j}\right)+ f\left( w^{n+1/2}_{i,j+1}\right) \right]
\end{array}
 \label{ntb08.01}
\end{equation}
and
\begin{equation}
\begin{array}{l}
 \displaystyle \int_{t_{n}}^{t_{n+1}} \int_{x_{i}}^{x_{i+1}} g(w(x,y_{j},t))dxdt \cong \\
 \\
 \displaystyle \cong \frac{\Delta x \Delta t}{2}\left[ g\left( w^{n+1/2}_{i,j}\right)+ g\left( w^{n+1/2}_{i+1,j}\right) \right],
\end{array}
 \label{ntb08.02}
\end{equation}
where
\begin{equation}
 w^{n+1/2}_{i,j} = \overline{w}^{n}_{i,j} - \frac{\Delta t}{2\Delta x}D_{x}f_{i,j} - \frac{\Delta t}{2\Delta y}D_{y}g_{i,j}\, , 
\end{equation}
with 
\[D_{x}f_{i,j}=MM\{\Delta_{x} f_{i+1/2,j},\Delta_{x} f_{i-1/2,j}\},\]
in which $\Delta_{x} f_{i+1/2,j}=f_{i+1,j}-f_{i,j}$ and \[D_{y}g_{i,j}=MM\{\Delta_{y} g_{i,j+1/2},\Delta_{y} g_{i,j-1/2}\},\] with $\Delta_{y} g_{i,j+1/2}=g_{i,j+1}-g_{i,j}$.

Replacing equations (\ref{ntb08}), (\ref{ntb08.01}) and (\ref{ntb08.02}) on equation (\ref{ntb06}), we get the \textbf{bidimensional Nessyahu-Tadmor method} on staggered mesh,

\begin{equation}
  \begin{array}{l}
 \displaystyle \overline{w}_{i+1/2,j+1/2}^{n+1}=  \displaystyle \frac{1}{4}\left(\overline{w}_{i,j}^{n}+\overline{w}_{i+1,j}^{n}+ \overline{w}_{i,j+1}^{n} +\overline{w}_{i+1,j+1}^{n}\right) +  \\
     \\
 \displaystyle + \frac{1}{16}\left[(D_{x}w_{i,j} - D_{x}w_{i+1,j}) + (D_{x}w_{i,j+1} - D_{x}w_{i+1,j+1} \right]\;+   \\
    \\
                                                 \\
    \displaystyle +\frac{1}{16}\left[(D_{y}w_{i,j} - D_{y}w_{i,j+1}) +  (D_{y}w_{i+1,j} - D_{y}w_{i+1,j+1})\right]\;+\\
                                                  \\
  \displaystyle - \frac{\Delta t}{2\Delta x}\left[ f\left( w^{n+1/2}_{i+1,j}\right)- f\left( w^{n+1/2}_{i,j}\right)\right] + \\
    \\
  \displaystyle  - \frac{\Delta t}{2\Delta x}\left[ f\left( w^{n+1/2}_{i+1,j+1}\right)- f\left( w^{n+1/2}_{i,j+1}\right)\right] + \\
                                                 \\
  \displaystyle - \frac{\Delta t}{2\Delta y}\left[ g\left( w^{n+1/2}_{i,j+1}\right)- g\left( w^{n+1/2}_{i,j}\right) \right] + \\
     \\
   \displaystyle - \frac{\Delta t}{2\Delta y}\left[ g\left( w^{n+1/2}_{i+1,j+1}\right)- g\left( w^{n+1/2}_{i+1,j}\right) \right]\, . \\
 \end{array}\;
 \label{ntb09}
\end{equation}

Hence, equation (\ref{ntb09}) is an approximation to equation (\ref{ntb06}). So, we can use equation (\ref{ntb09}) for evolution of $w$.

If the $w$ function were piecewise constant we would have 
$$D_{x}w=D_{y}w=D_{x}f=D_{y}g=0,$$ 
and then the equation (\ref{ntb09}) would reduce to \textbf{bidimensional Lax-Friedrichs method} on staggered mesh. These methods may also be obtained for a non-staggered mesh \citep{leveque2002,nessyahu,thomas}.


\section{Jets simulations}
\subsection{Computational support}

We executed our code in a simple personal computer with the following configuration: intel core i7 processor (2.9 GHz), quad core and 6 Gb RAM memory DDR 3; Linux/GNU 64 bits, kernel 3.16.0-30 and 5786.81 bogomaps (cpuinfo terminal command). We used gfortran compiler version 4.9.1.

\subsection{Initial data}

We made the following initial data to build a favourable setting for studying the formation of jets and its substructures: spatial domain was delimited in $[1.1\;\; 20]\times [0,\pi /2]$, that is $r\in[1.1\;\; 20]$ and $\theta\in [0,\pi /2]$; we used  $\delta = 0.125$, that is, $\theta_D=\tan^{-1}\delta$ for disk aperture and $r_{D}=3r_{S}$, three times the BH radius, for inner edge of disk. We chose $\Gamma=5/3$ and $H=1.3$ for specific heat and enthalpy, respectively.

\subsection{First scenario}\label{first}

\begin{figure}
 \includegraphics[scale=0.5]{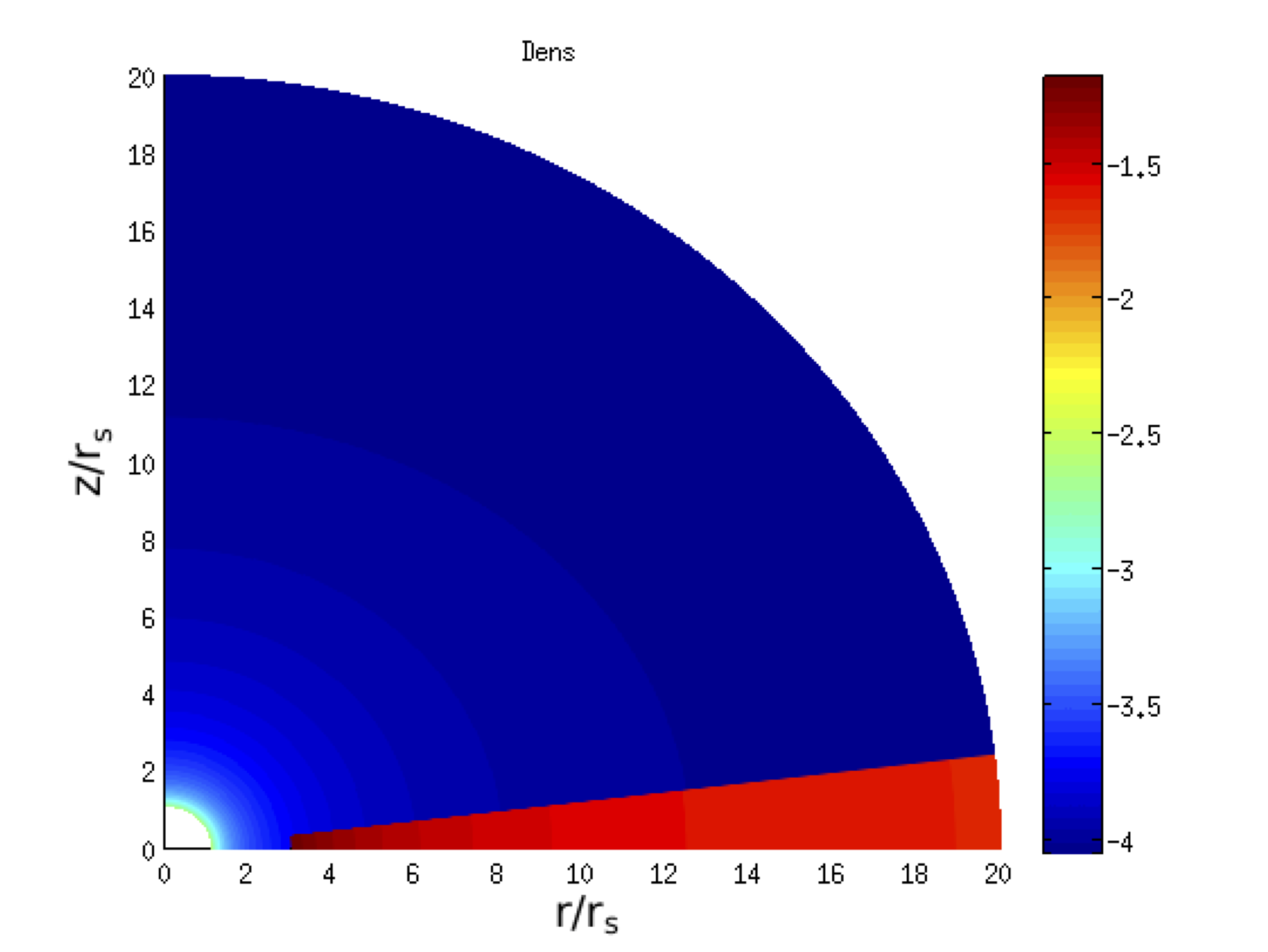}
 \caption{{\small The initial profile of the density.}}
 \label{jat02in01.0}
\end{figure}

\begin{figure}
 \centering
 \includegraphics[scale=0.5]{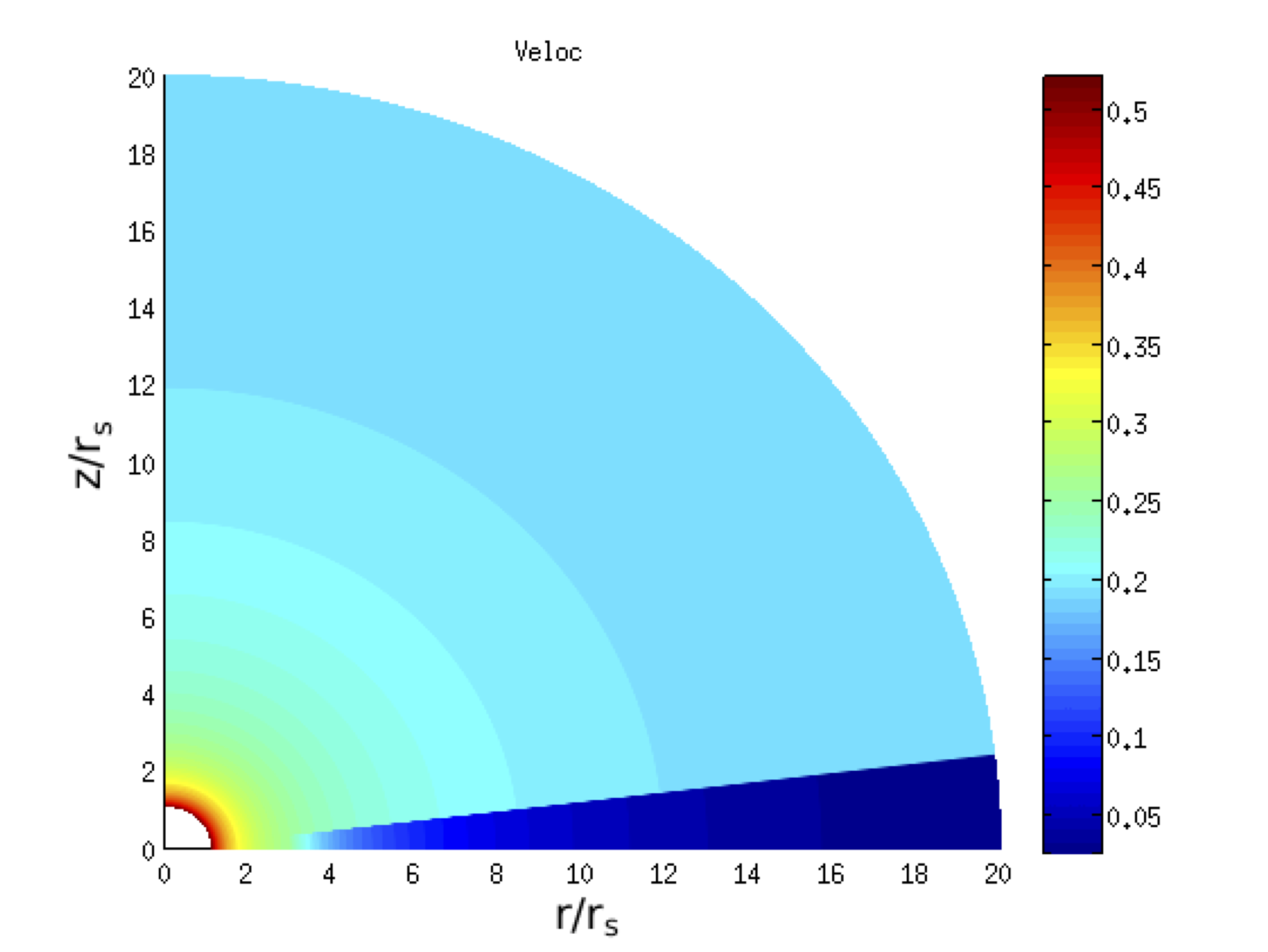}
 \caption{\small The initial profile of the magnitude of the velocity.}
 \label{jat02in01.1}
\end{figure}

\begin{figure}
\centering
 \includegraphics[scale=0.5]{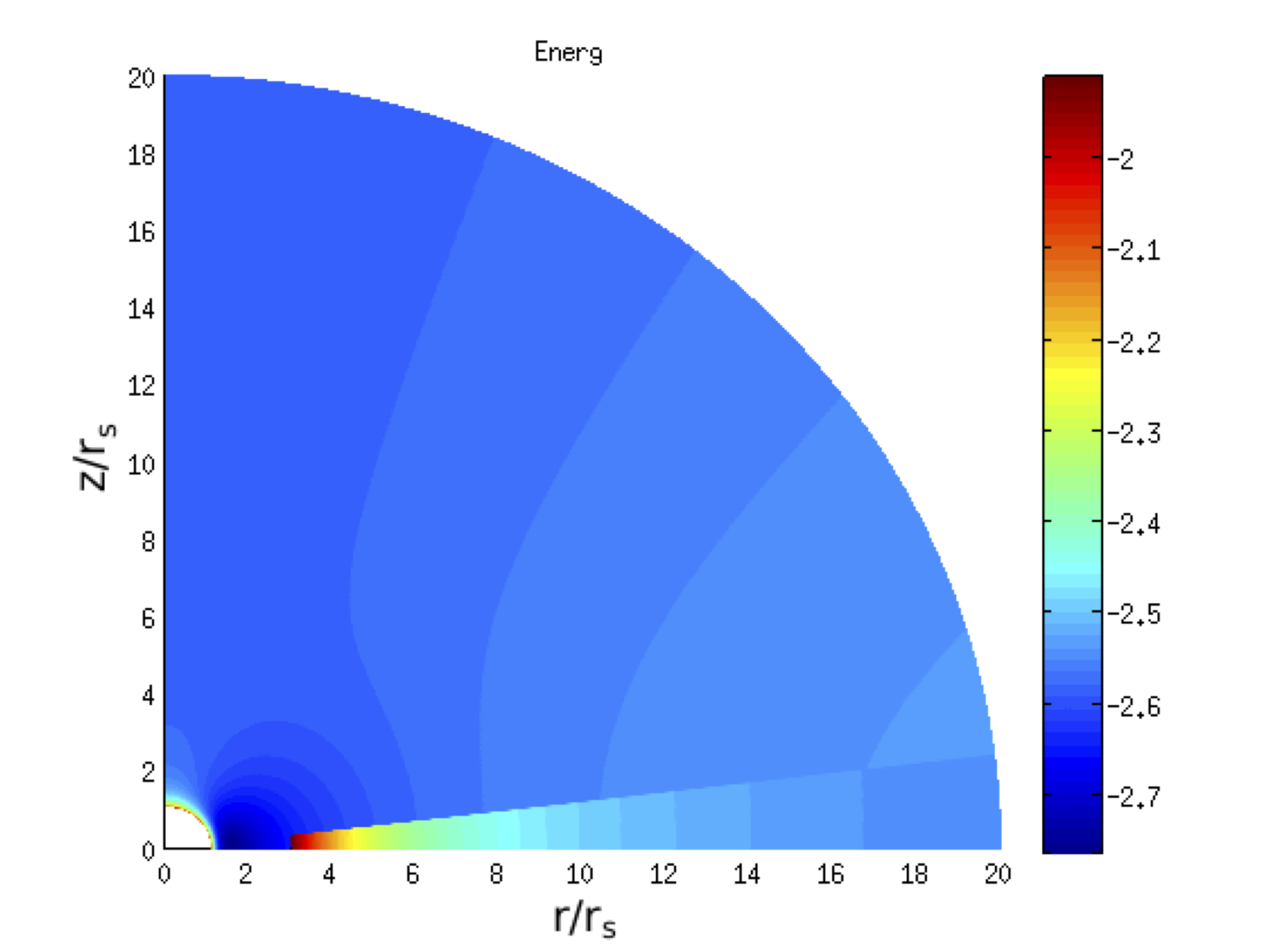}
 \caption{\small The initial profile of the total energy.}
 \label{jat02in02.1}
\end{figure}

\begin{figure}
 \centering
 \includegraphics[scale=0.5]{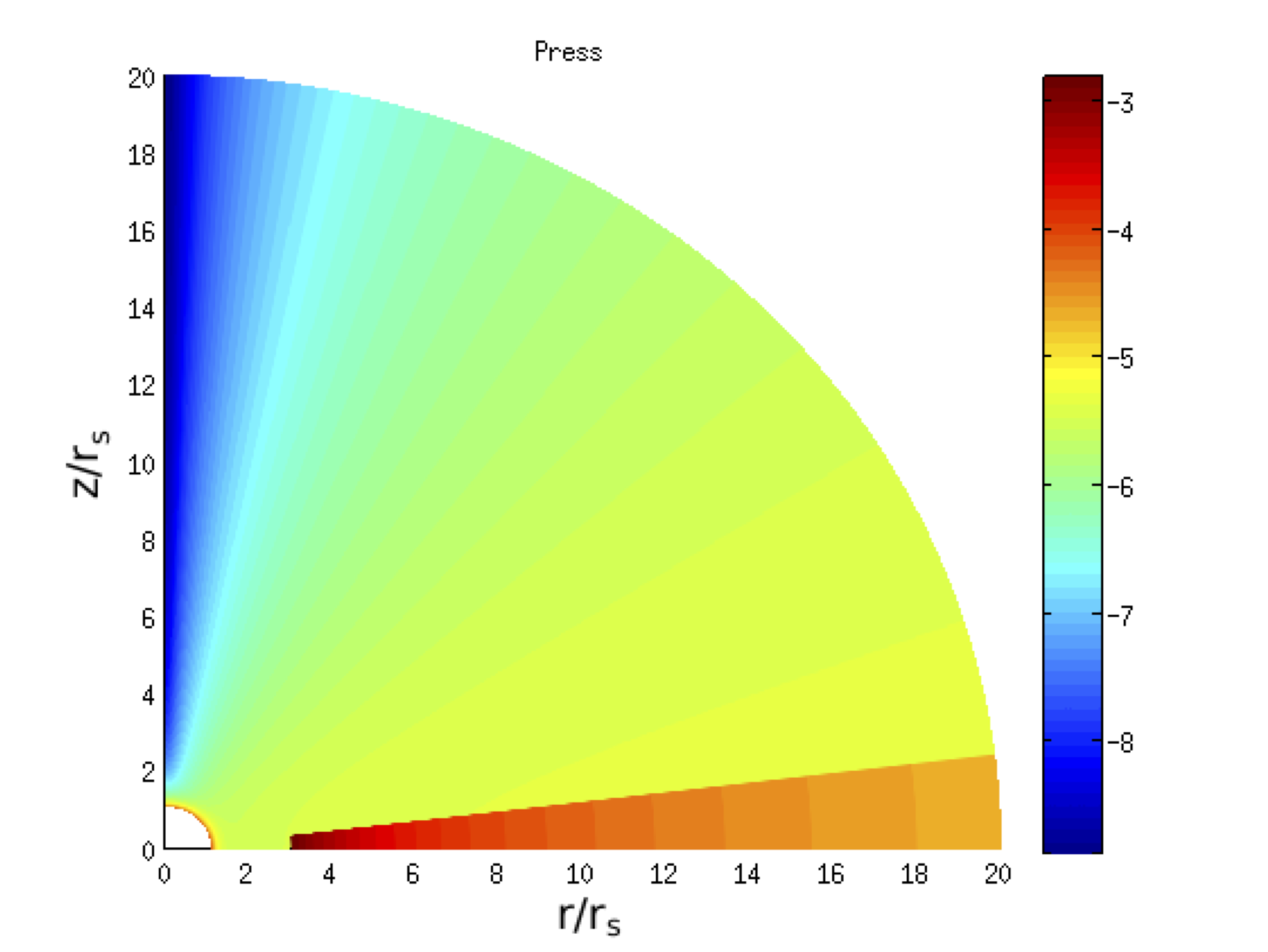}
 \caption{\small Initial profile of the magnitude of the pressure.}
 \label{jat02in02.2}
\end{figure}

\begin{figure}
 \centering
 \includegraphics[scale=0.5]{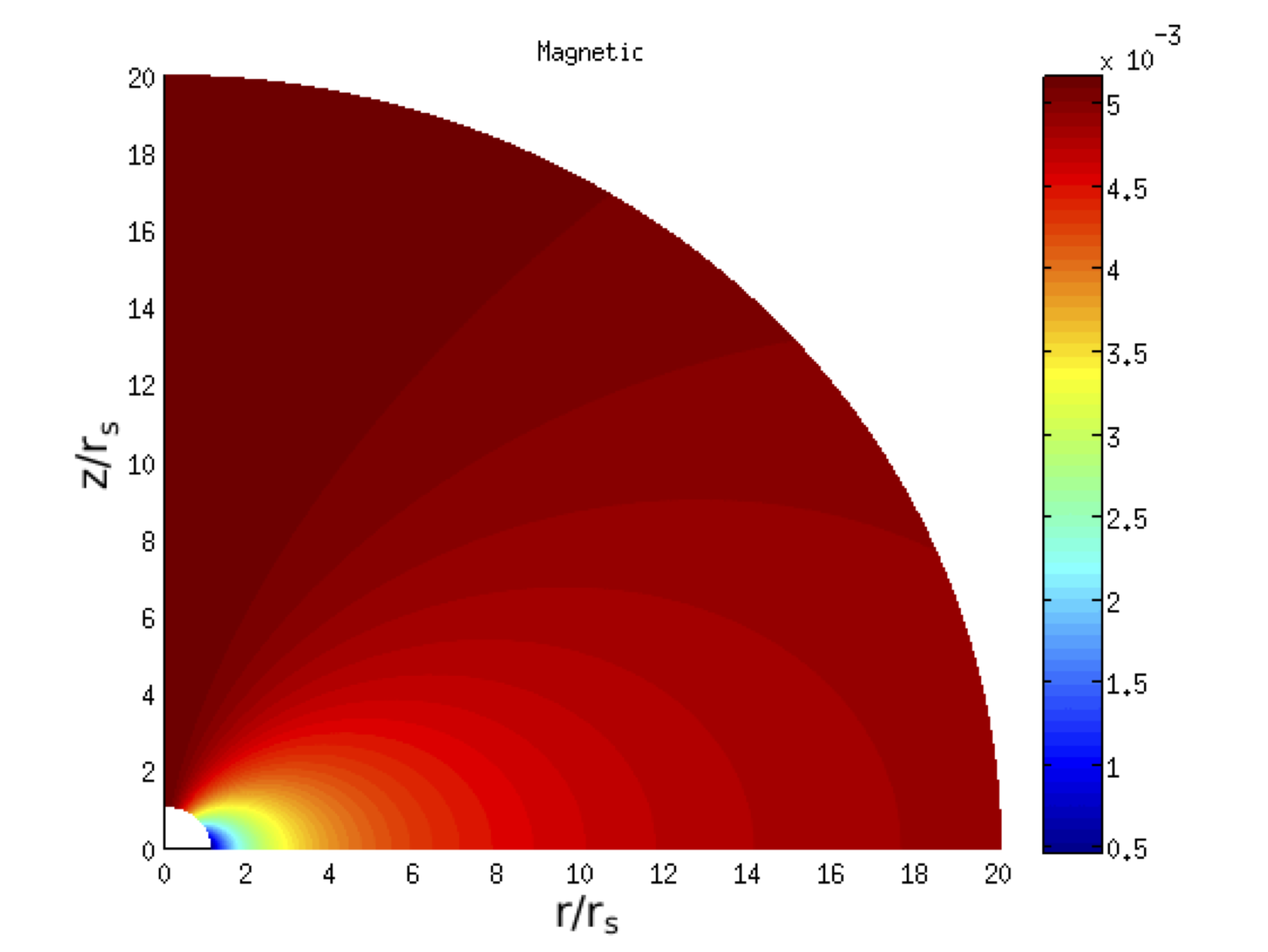}
 \caption{\small Initial profile of the magnitude of the magnetic field.}
 \label{jat02in03}
\end{figure}

In this case, we considered a disk $10,000$ times denser than magnetosphere ($\kappa_{\rho}=10^{4}$). Thus, we have initial conditions to density (Fig. \ref{jat02in01.0}), magnitude of velocity (Fig. \ref{jat02in01.1}), total energy (Fig. \ref{jat02in02.1}), pressure (Fig. \ref{jat02in02.2}) and magnitude of magnetic field (Fig. \ref{jat02in03}). The graphs of the values of density, pressure and total energy were plotted in logarithmic scale to base 10. To graphs we used $\log_{10}\left( |D|^{2} \right)$, $\log_{10}\left( |\mathcal{E}|^{2} \right)$, $\log_{10}\left( |p|^{2} \right)$ and $\log_{10}\left( |\textbf{B}|^{2} \right)$ whose axis are $r/r_{S}$ and $z/r_{S}$.

 We performed this initial conditions in a $630\times 630$ grid, that is each interval $[1.1\; , 20]$ and $[0\; ,\pi /2]$ subdivided in $630^2$ cells.
 
\begin{figure}
 \centering
 \includegraphics[scale=0.5]{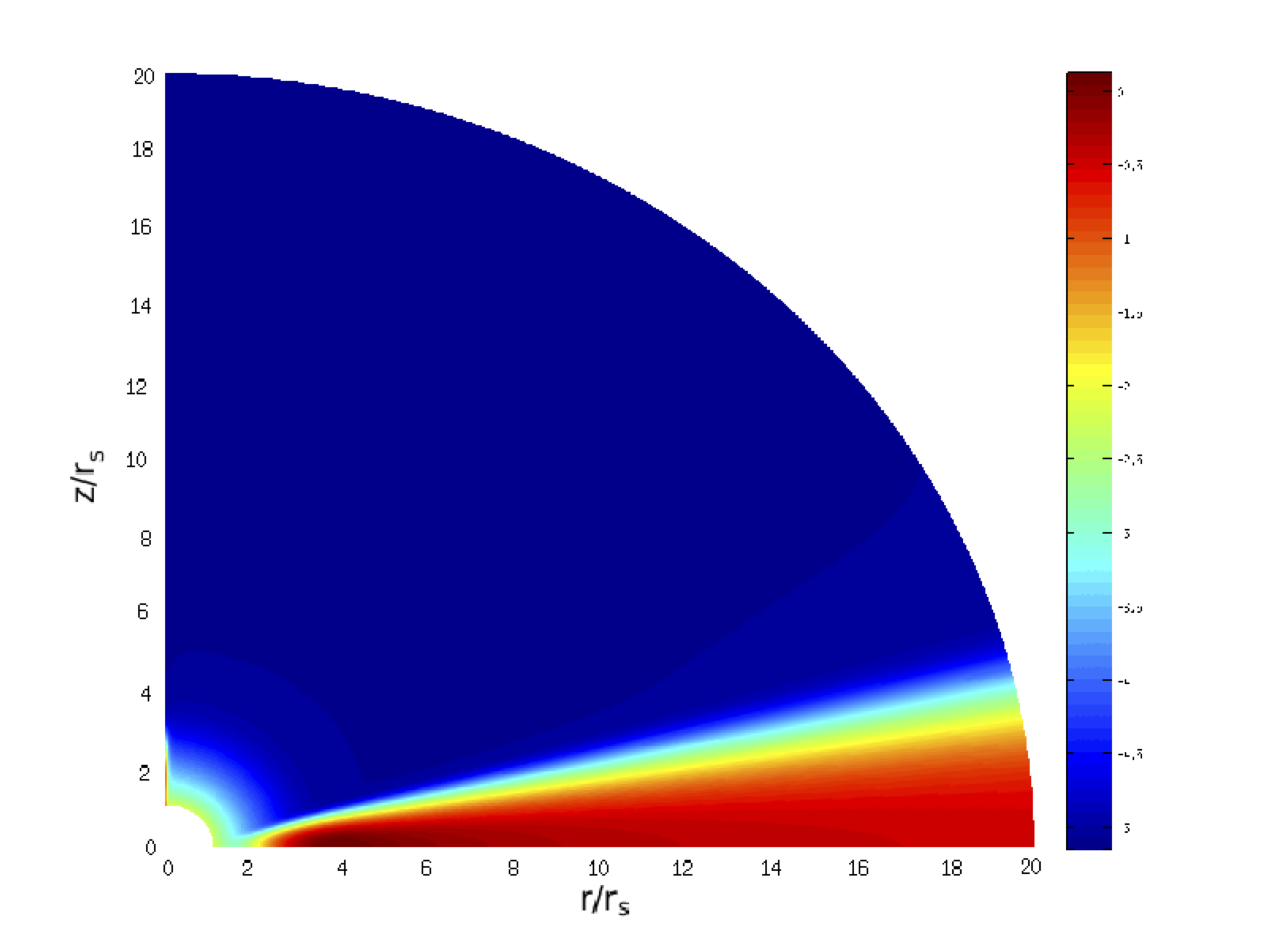}
 \caption{\small Time evolution: density profile in logarithmic scale after 810 iterations.}
 \label{00027}
\end{figure}

\begin{figure}
 \centering
 \includegraphics[scale=0.5]{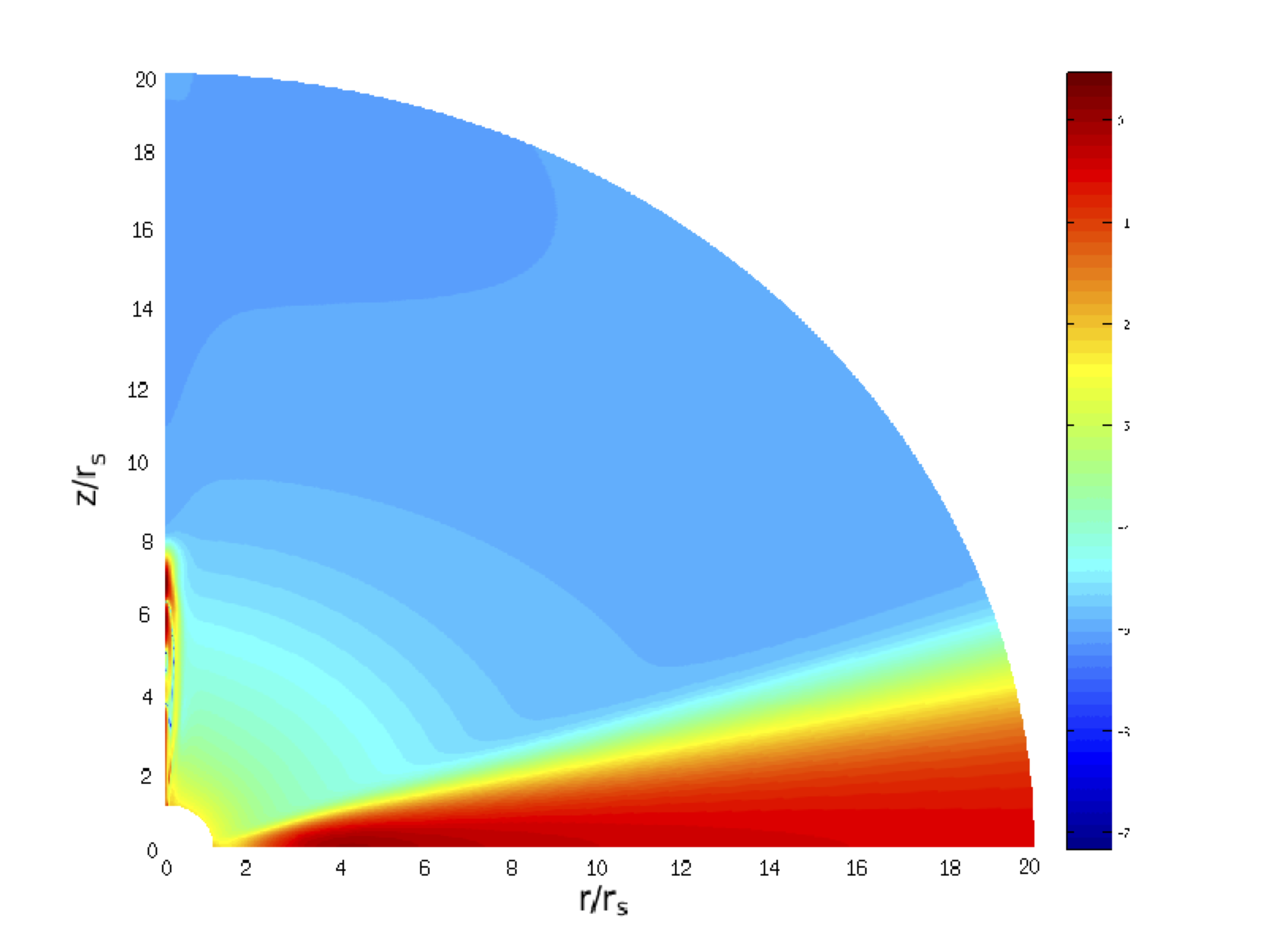}
 \caption{\small Time evolution: density profile in logarithmic scale after 1,500 iterations.}
 \label{00050}
\end{figure}

\begin{figure}
 \centering
 \includegraphics[scale=0.5]{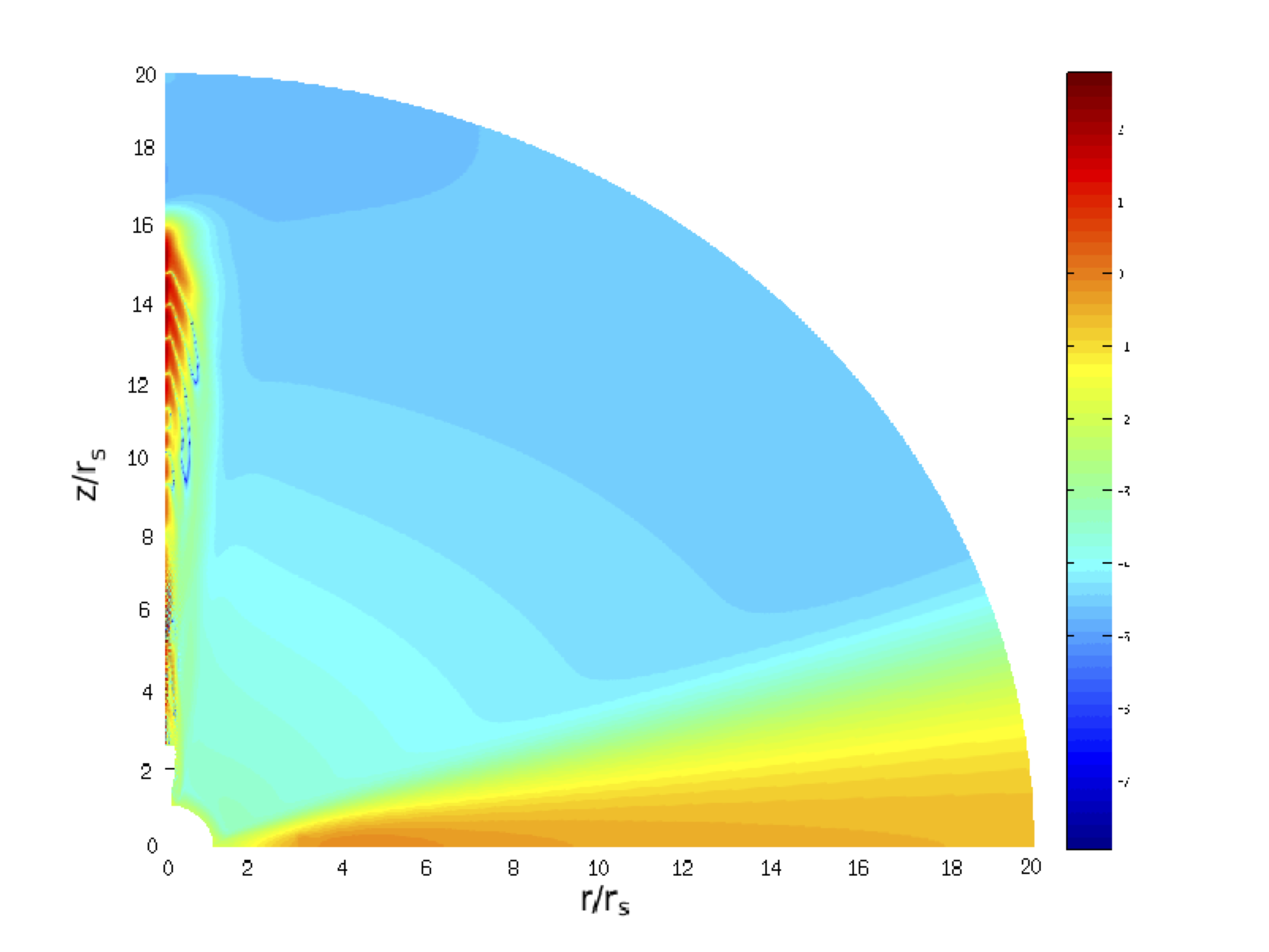}
 \caption{\small Time evolution: density profile in logarithmic scale after 2,200 iterations.}
 \label{00074}
 \end{figure}

 In Fig. \ref{00027} one can see the accretion disk is attracted by central BH while magnetosphere falls, thus the matter accumulates in the vicinity of BH and a jet is being ejected at the BH pole. The jet goes on its evolution, develops its substructures and reaches distance of sixteen times the Schwarzschild radii -- see Figs \ref{00050} and \ref{00074}.
 
In this simulation, the spatial and temporal increments were \[ \Delta t = 0.005 \min\{\Delta r, \Delta\theta \}, \] where $\Delta r = \frac{18.9}{630}$ and $\Delta\theta=\frac{\pi}{1260}$.
 
From the initial conditions to final time there were 2,220 iterations, which is about $11 R_s $ coordinate time lapse. Up to this time evolution we have obtained good approximate non oscillatory solutions. 

\begin{figure}
 \centering
 \includegraphics[scale=0.6]{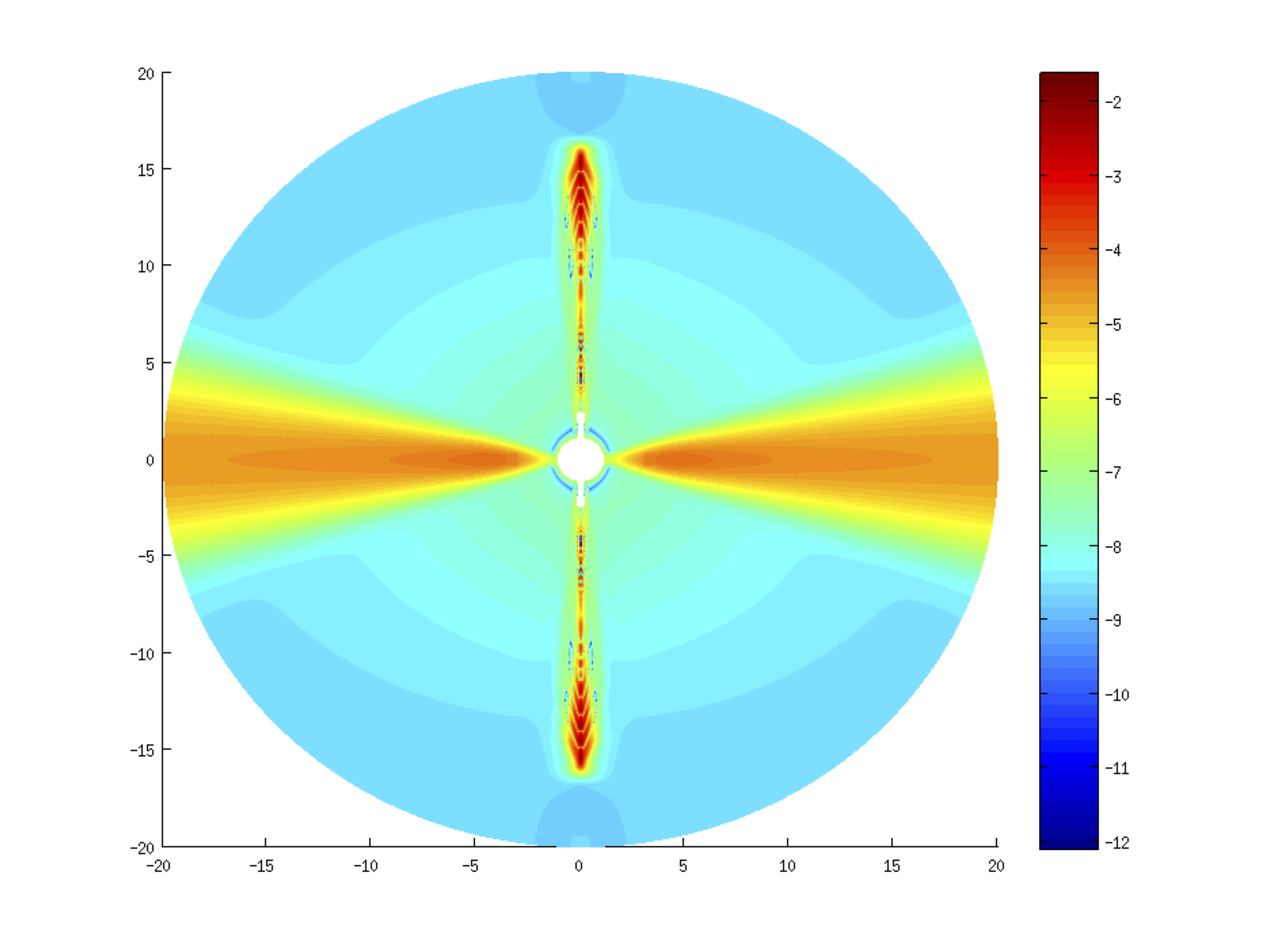}
 \caption{\small Graph of density in full domain $\theta \in [0,2\pi]$. Note that we got two jets going out the poles and a disk surrounding the central BH.}
 \label{ex1_0074t}
\end{figure}

\begin{figure}
 \centering
 \includegraphics[scale=0.6]{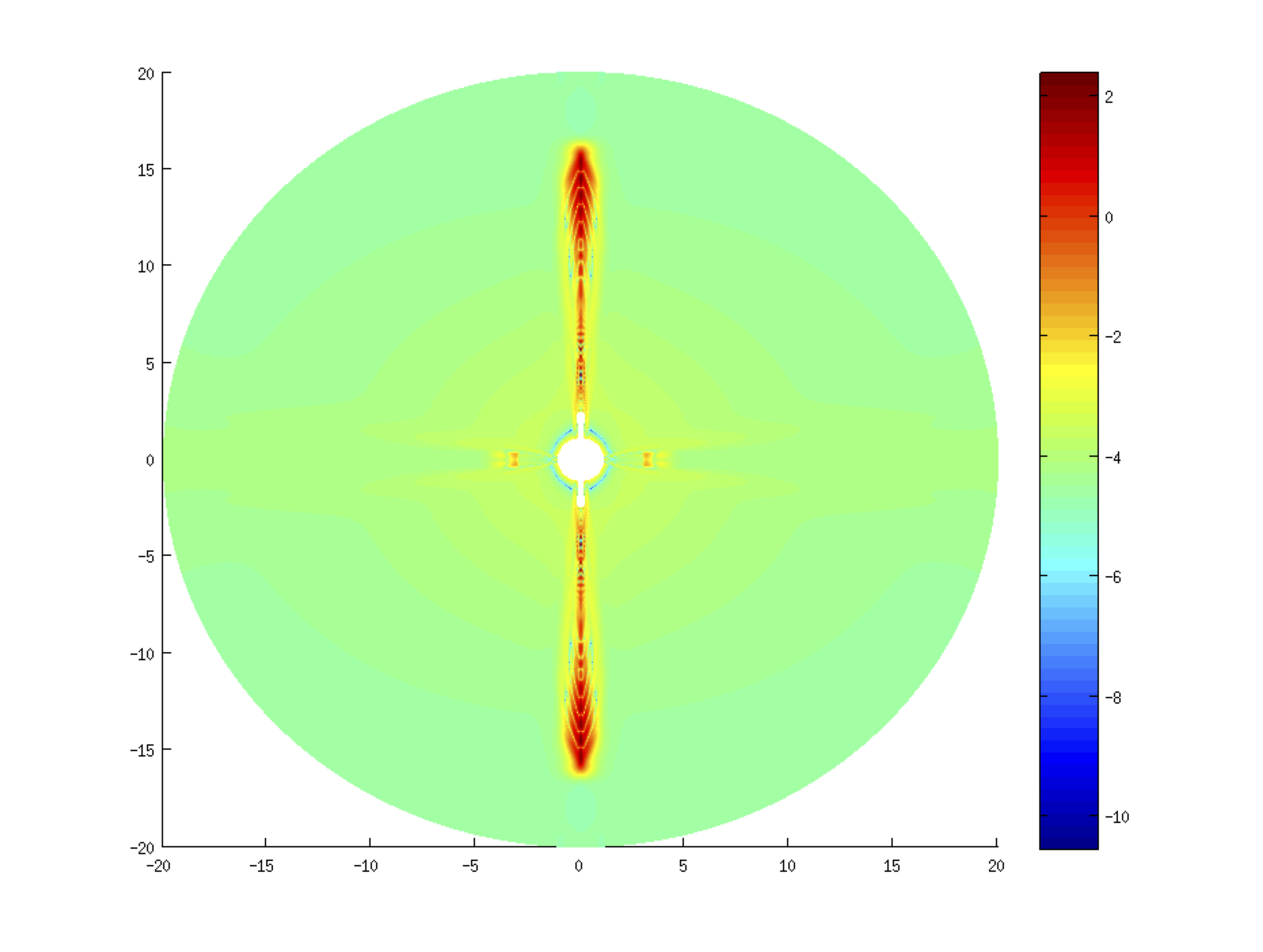}
 \caption{\small Graph of total energy in full domain $\theta \in [0,2\pi]$.}
 \label{ex1_0074}
\end{figure}

 All theses figures were plotted on the first quarter. By axial symmetry, we can plot the numerical solution into $[0,\pi /2]$. Thus, for density quantity we got the graphs Fig. \ref{ex1_0074t} and Fig. \ref{ex1_0074} represents the total energy profile. Therefore, we have a disk around the BH and symmetrical jets coming out of poles. 
 
\begin{figure}
 \centering
 \includegraphics[scale=0.6]{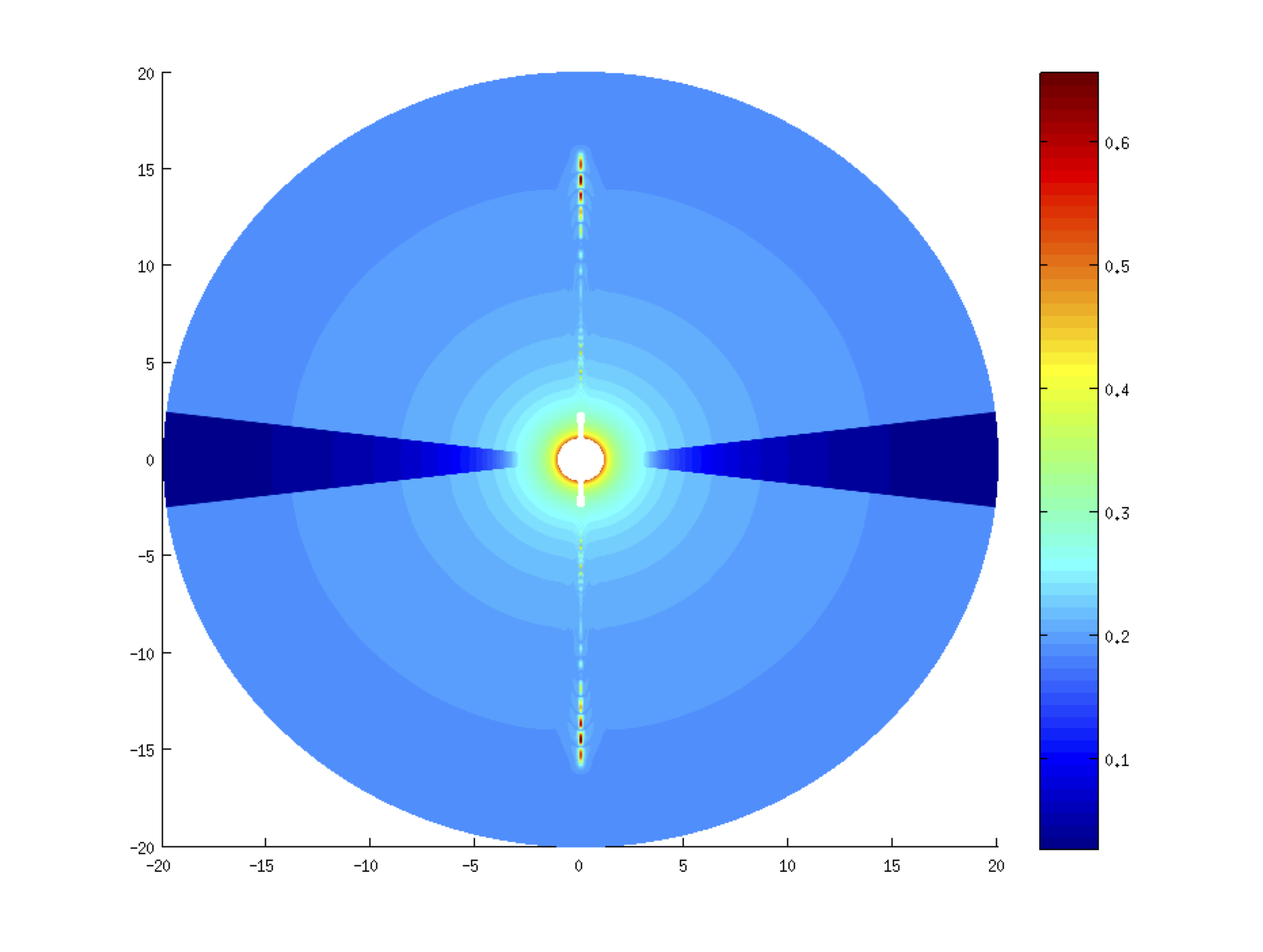}
 \caption{\small Graph of the magnitude of velocity after 2,200 iterations. Note that jet assumes relativistic values, see colourbar}
 \label{00074veloc}
\end{figure}

In this period, we we got relativistic jets because the magnitude velocity reached relativistic values. See Fig. \ref{00074veloc}. This simulation reveals some substructures inside the jet and the magnitude velocity profile accompanies time evolution of density and of the total energy.

\subsection{Other meshes}

The same example of subsection \ref{first} was executed into other grids to evaluate numerically the approximate solutions. We used $210\times 210$ and $840\times 840$ grids. Practically, numerical solutions provided by $840\times 840$  showed no difference when compared with solutions obtained into $630\times 630$ grid.

\begin{figure}
 \centering
 \includegraphics[scale=0.6]{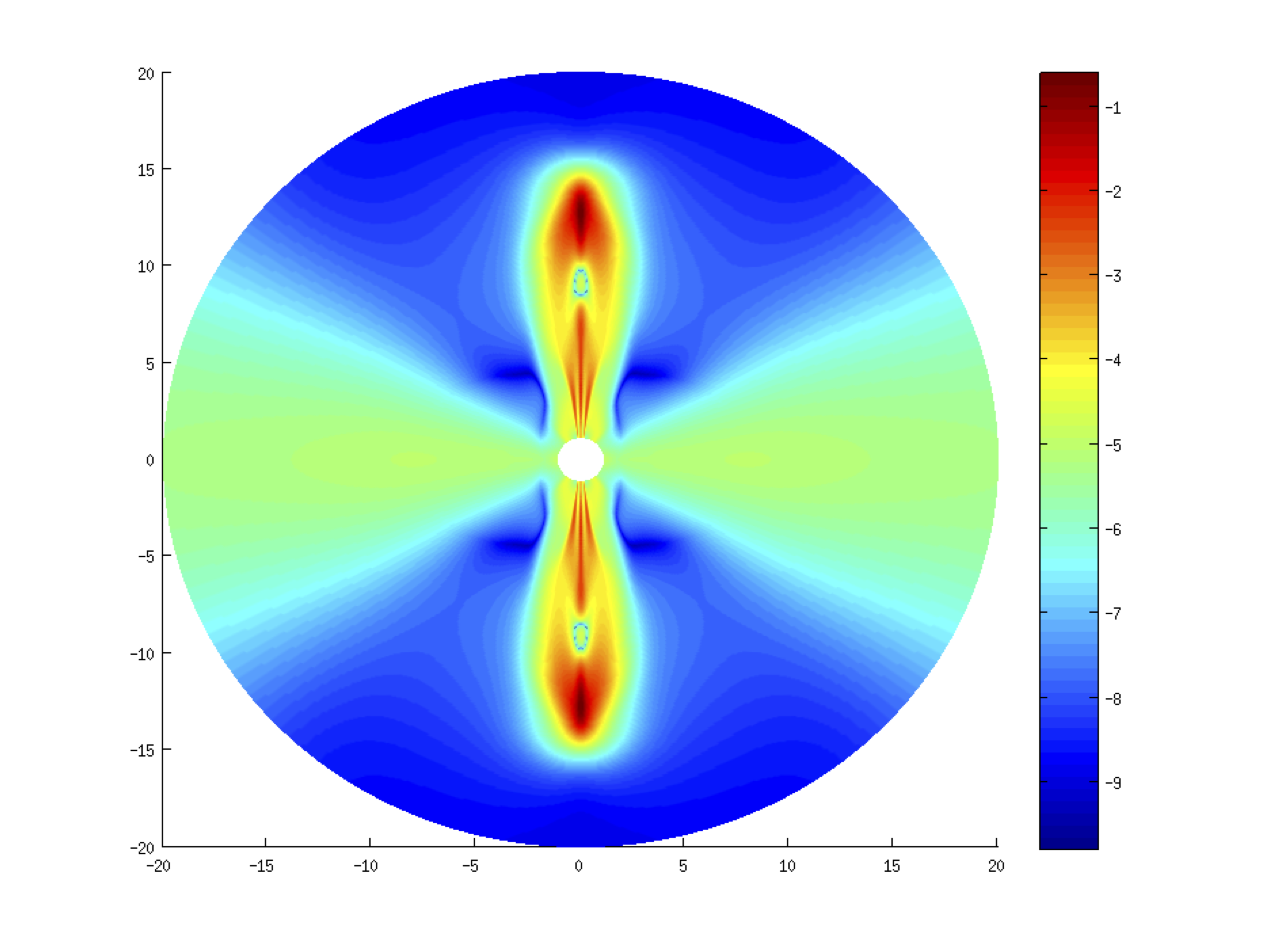}
 \caption{\small Graph of density in a $210\times 210$ mesh after 2,200 iterations}
 \label{ex3_0074d}
\end{figure}

\begin{figure}
 \centering
 \includegraphics[scale=0.6]{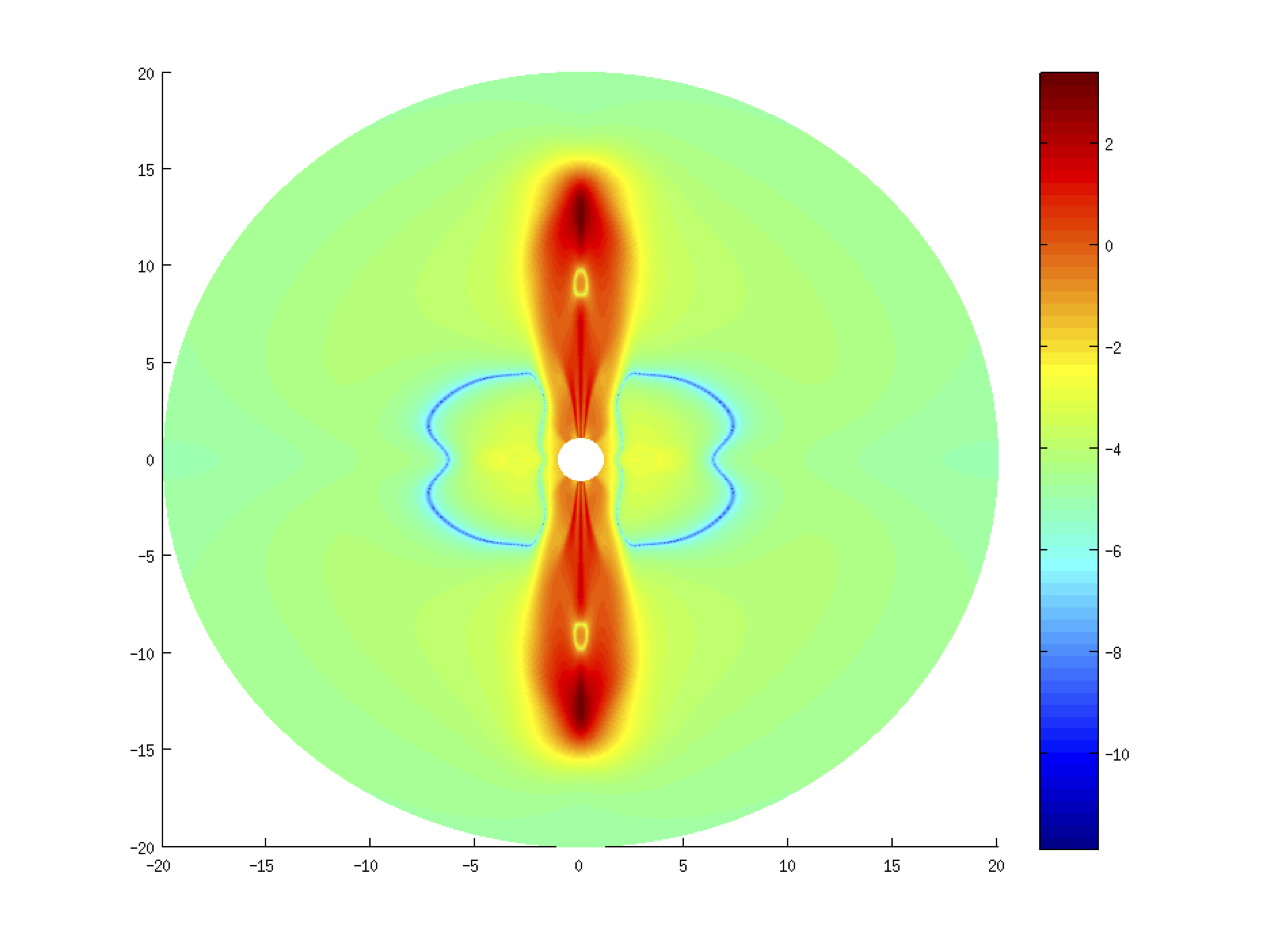}
 \caption{\small Graph of total energy in a $210\times 210$ mesh after 2,200 iterations.}
 \label{ex3_0074e}
\end{figure}

With coarse grid we got a jet of low resolution, see Fig. \ref{ex3_0074d} and \ref{ex3_0074e}. In this situation, we see some substructures but we don't have details of internal structures jets. However, this simulation is able to check trends of numerical solutions and expose some features of numerical method and the phenomenon under study. 



Considering the same settings for a different mesh, we have a large difference between the meshes with $210^2$ cells compared to the $630^2$ one. But there is not significant difference between $630^2$ grid and $840^2$ one. So our numerical solution is closed to the analytical solution when grid is sufficiently refined.

\begin{figure}
 \centering
 \includegraphics[scale=0.6]{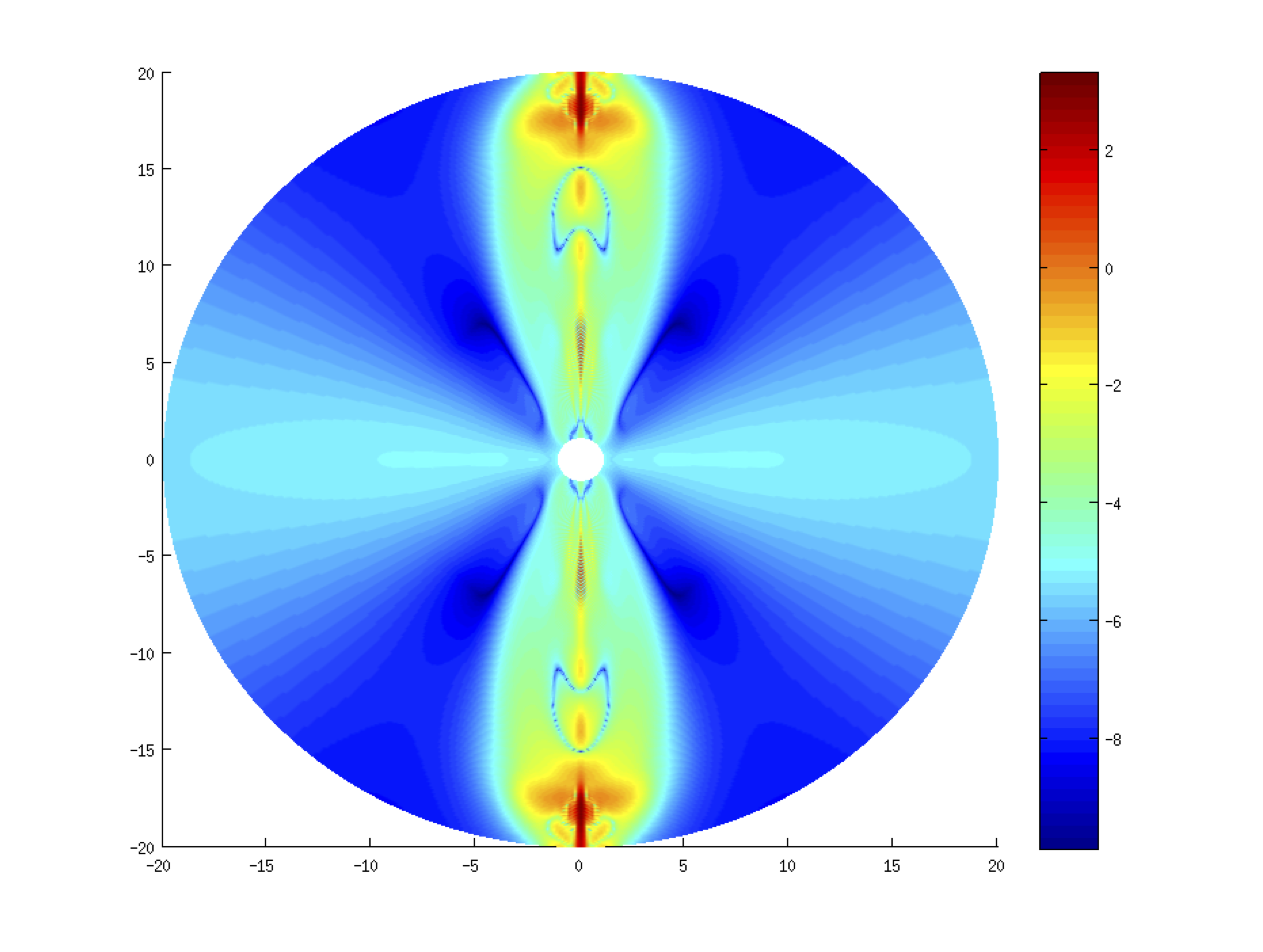}
 \caption{\small Graph of density in a $210\times 210$ mesh after 2,910 iterations. Here we have jets exceeding the length of $20r_{S}$.}
 \label{ex3_2910d}
\end{figure}

\begin{figure}
 \centering
 \includegraphics[scale=0.6]{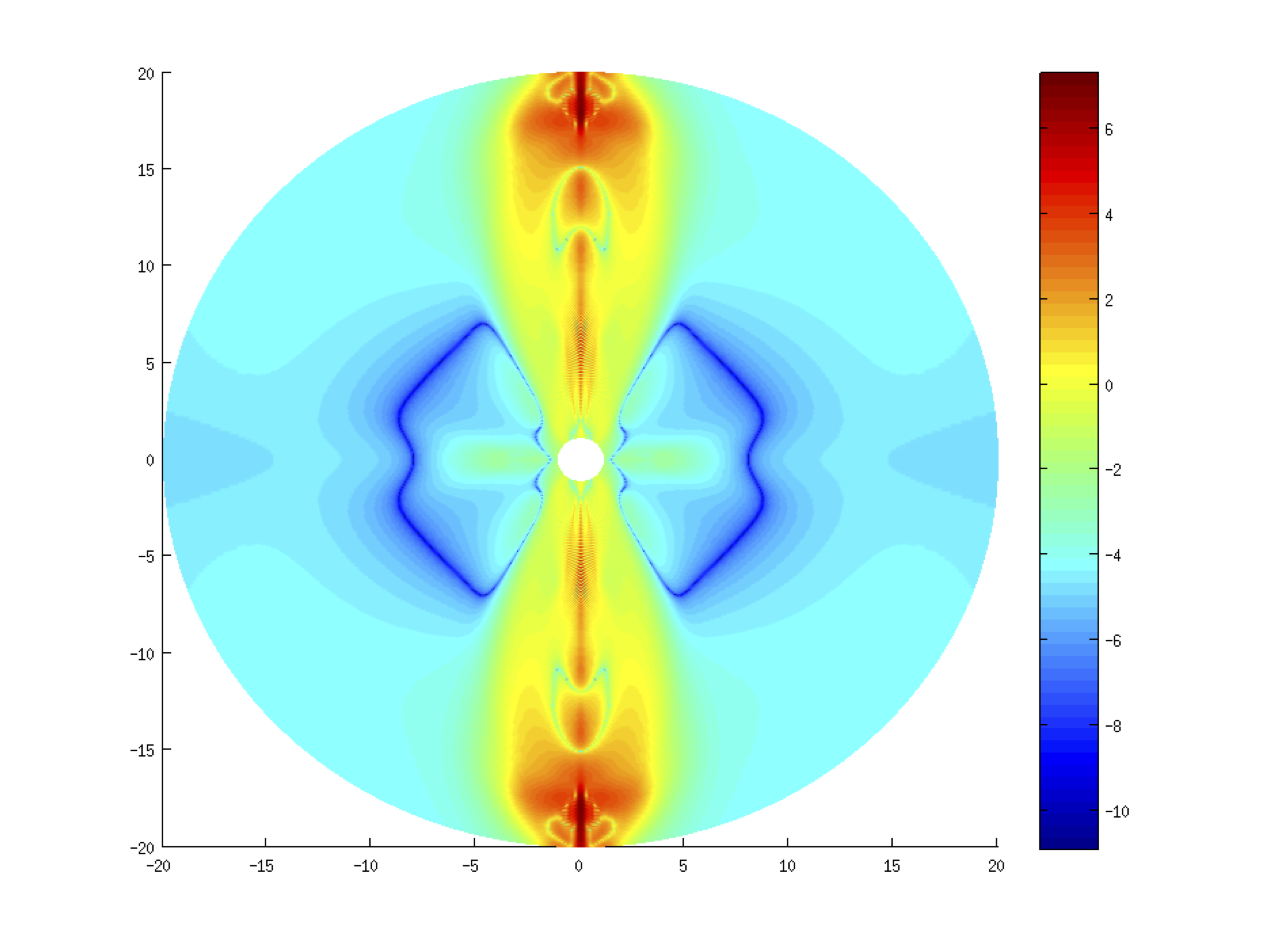}
 \caption{\small Graph of total energy in a $210\times 210$ mesh after 2,910 iterations.}
 \label{ex3_2910e}
\end{figure}

To continue the evolution in the mesh $210^2$, the jet exceeds the distance of $20 r_{S}$ (Figs. \ref{ex3_2910d} and \ref{ex3_2910e}). In this case one notice that the jet undergoes for a narrowing and a possible substructure begins to be formed. These observations have meant that we have created a new scenario able to develop the substructure within the spatial domain.

In coarse grid, the relationship between spatial and temporal increments was \[ \Delta t = 0.0008 \min\{\Delta r, \Delta\theta \}, \] where $\Delta r = \frac{18.9}{210}$ and $\Delta\theta=\frac{\pi}{420}$.

\subsection{Second scenario}

In this case, we considered a disk $10$ times denser than first scenario (subsection \ref{first}), that is, we are a disk $100,000$ times denser than magnetosphere ($\kappa_{\rho}=10^{5}$). Its time evolution is similar to the previous case, but with these new settings, we observed the development of the substructure jet within the domain.

\begin{figure}
 \centering
 \includegraphics[scale=0.6]{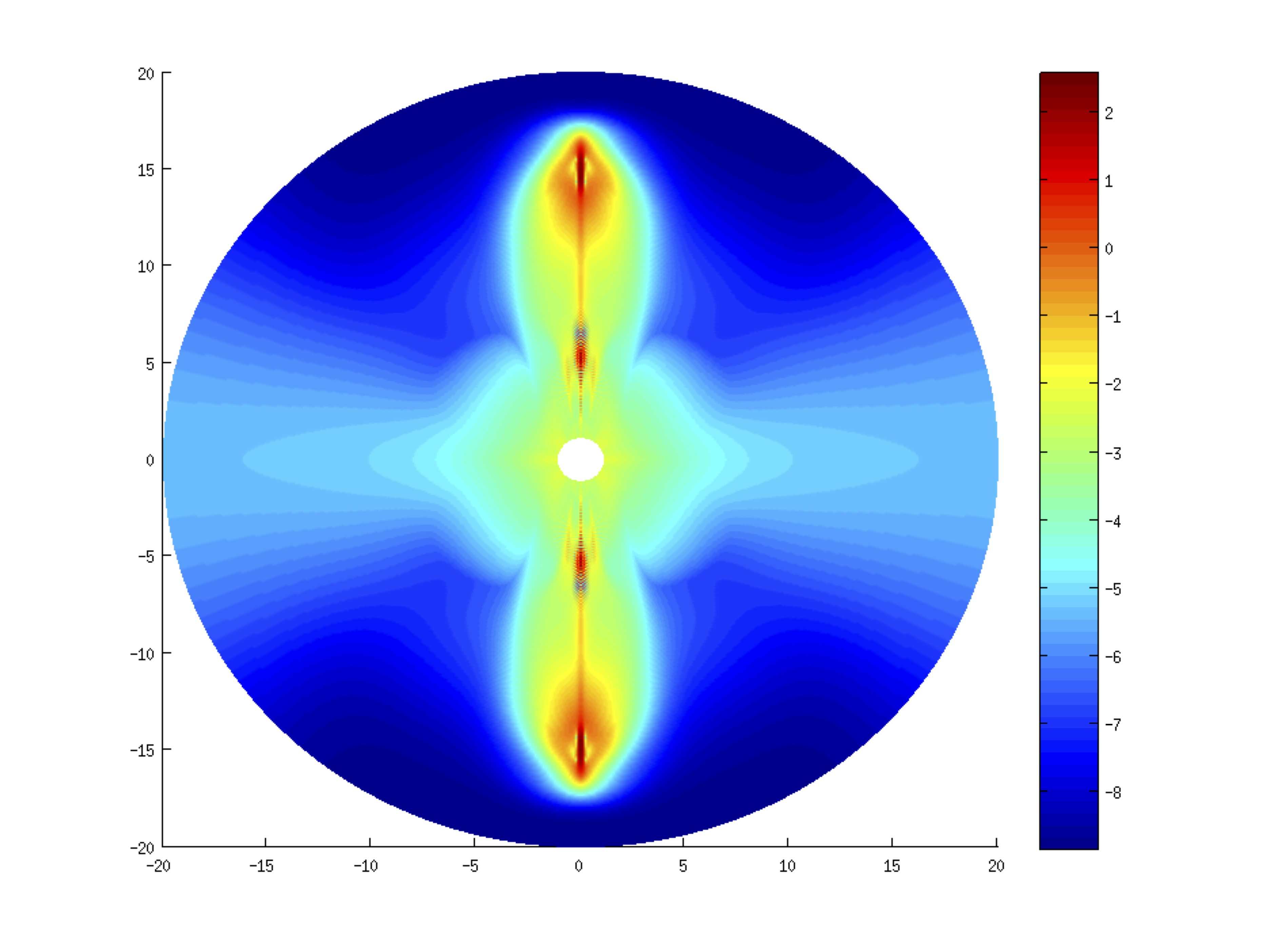}
 \caption{\small Graph of density after 2,340 iterations. In this moment, we got a jet before of its narrowing.}
 \label{ex2_2340d}
\end{figure}

\begin{figure}
 \centering
 \includegraphics[scale=0.6]{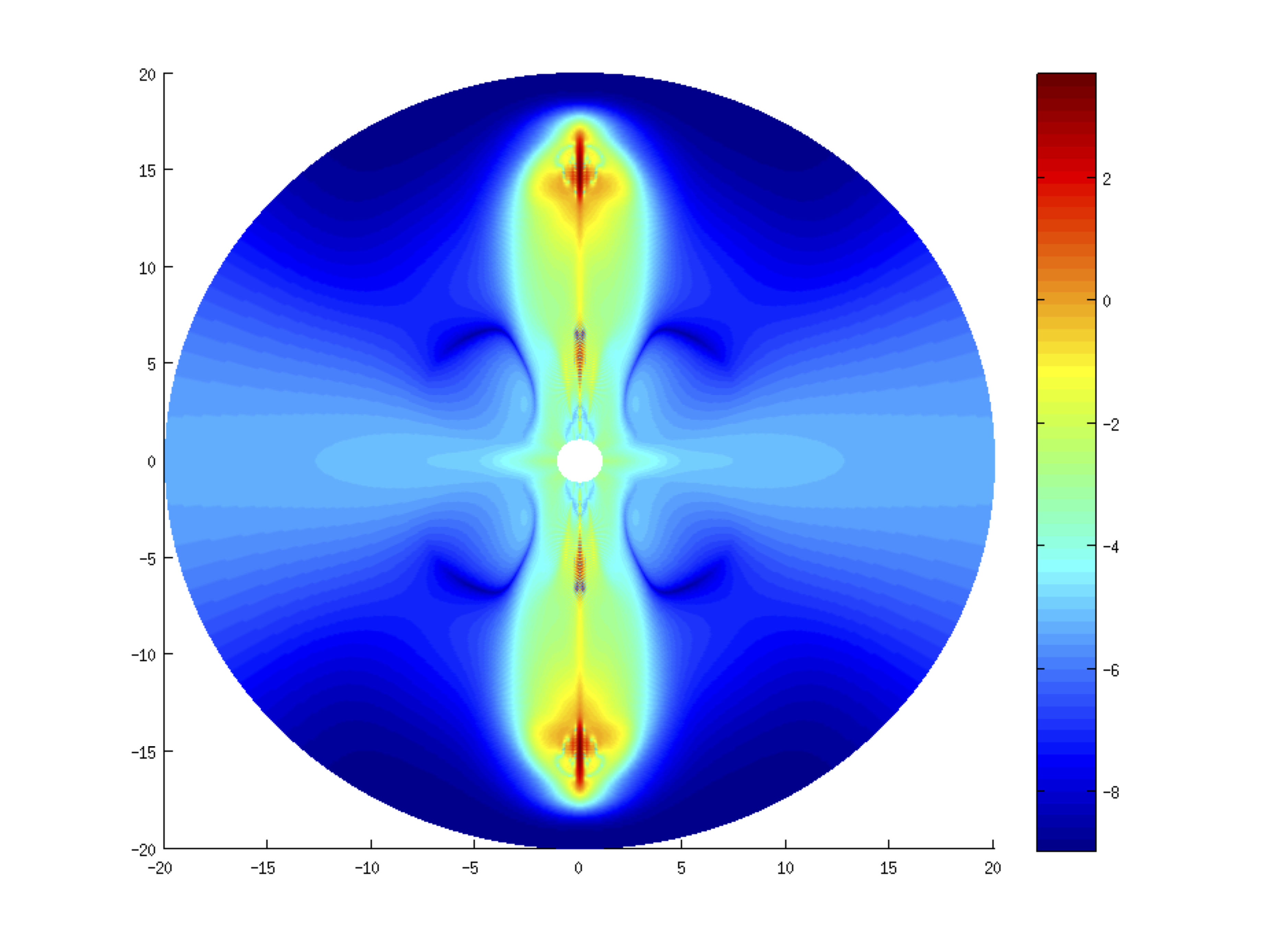}
 \caption{\small Density. This graph has development of a substructure of jet.}
 \label{ex2_2370d}
\end{figure}

\begin{figure}
 \centering
 \includegraphics[scale=0.6]{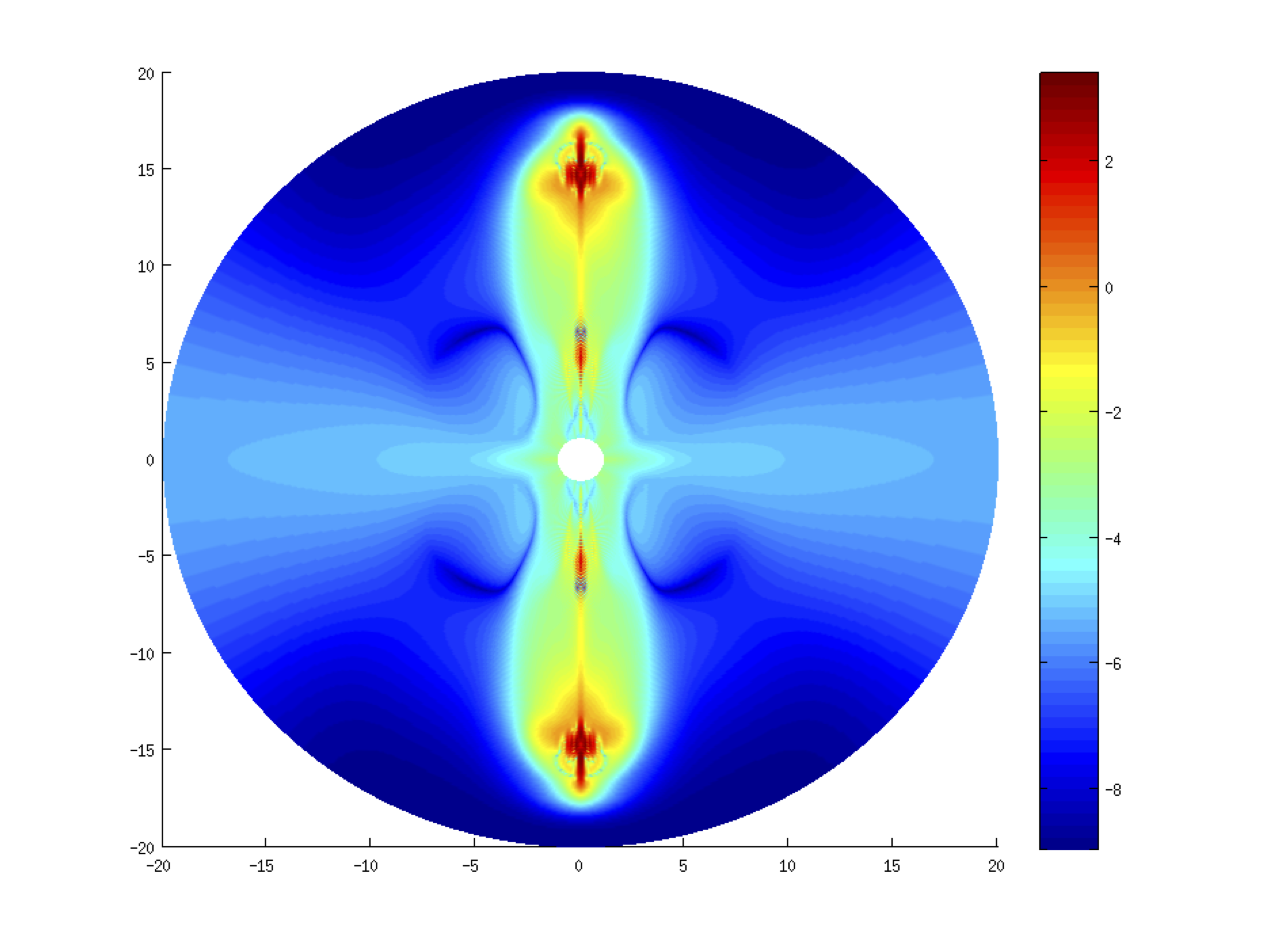}
 \caption{\small One lobe begins to stand out inside the tip of the jet.}
 \label{ex2_2373d}
\end{figure}

\begin{figure}
 \centering
 \includegraphics[scale=0.6]{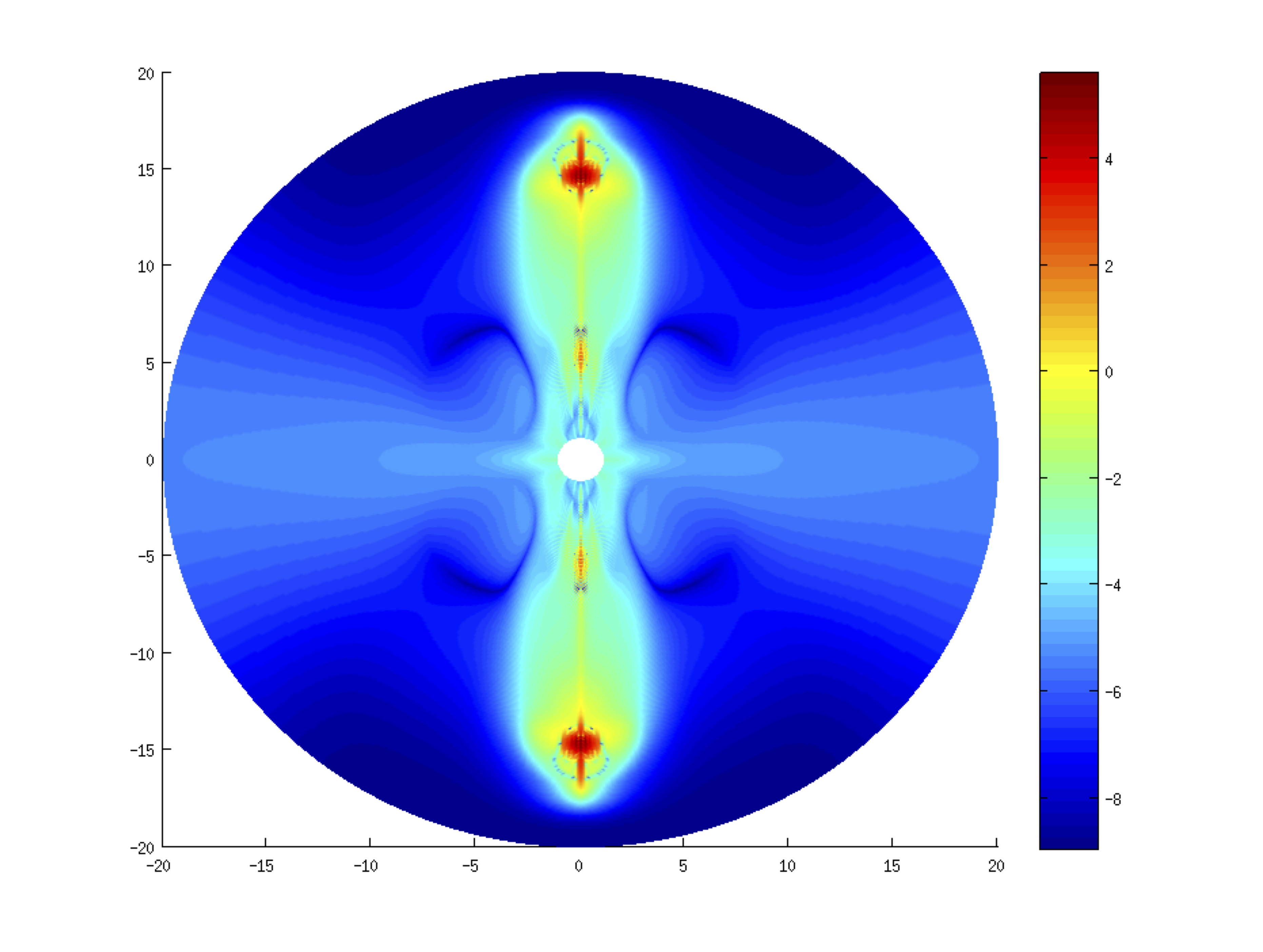}
 \caption{\small The lobe grows and stands out from other parts of the jet.}
 \label{ex2_2375d}
\end{figure}

\begin{figure}
 \centering
 \includegraphics[scale=0.6]{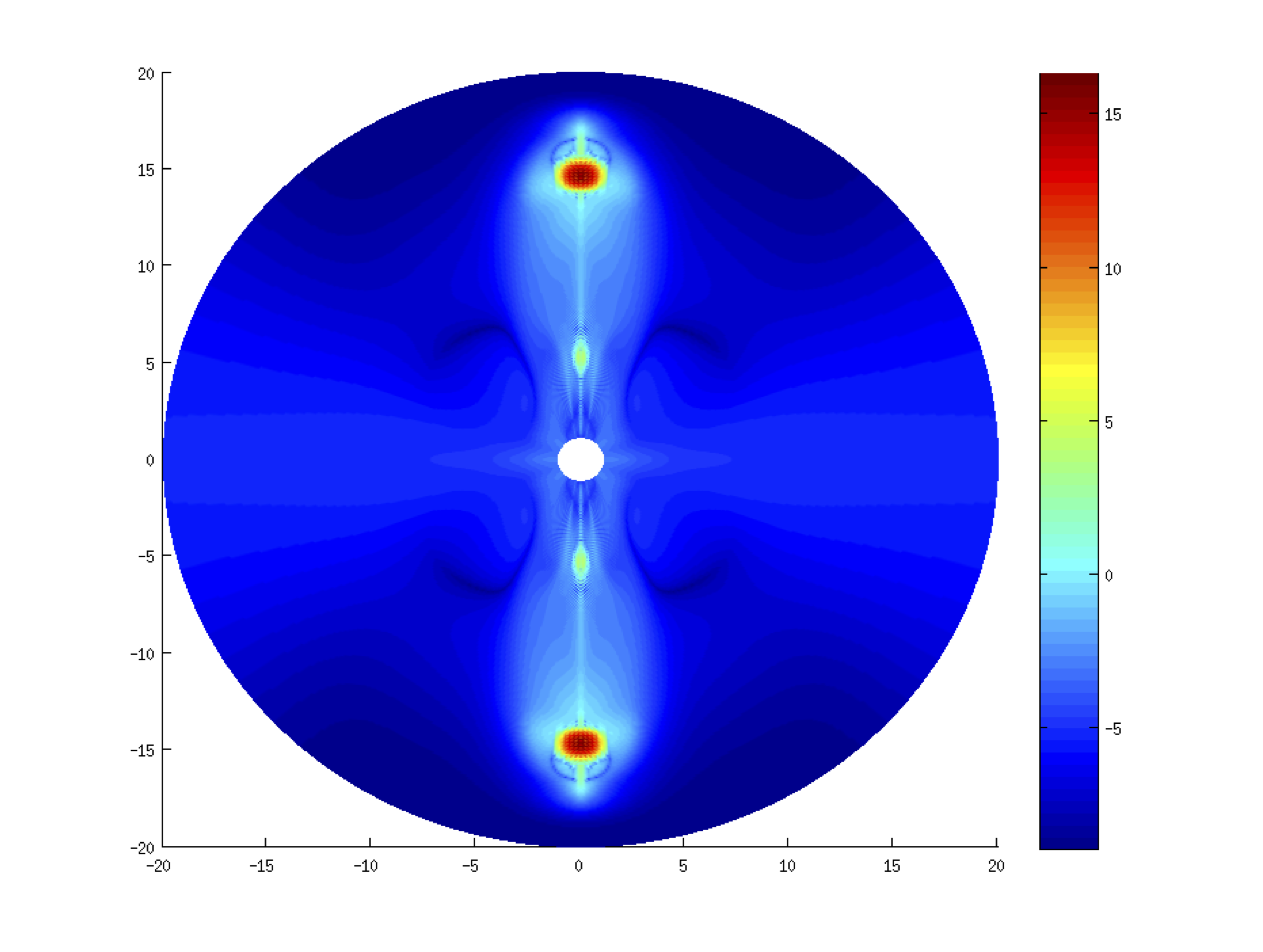}
 \caption{\small After 2378 iterations, the shape of the lobe is settled.}
 \label{ex2_2378d}
\end{figure}

In Fig. \ref{ex2_2340d} we have the jet formed with a length of $16r_{S}$. From that moment, the jet undergoes narrowing and a substructure is beginning (Fig. \ref{ex2_2370d}). By Fig. \ref{ex2_2373d} it is noted that there is an accumulation of material at $15r_{S}$. Such accumulation is developed and its shape is analogous to a Lobe (Fig. \ref{ex2_2375d}) inside of jet. After, the lobe grows and becomes denser (Fig. \ref{ex2_2378d}). So far, we had stable solutions and from that moment, the numerical solutions are unstable.

Because the jet is moving with a much higher speed than the magnetosphere, shocks form at its tip and its deceleration concentrates the matter in the form of a lobe (Figs. \ref{ex2_2340d}, \ref{ex2_2370d}, \ref{ex2_2373d}, \ref{ex2_2375d} and \ref{ex2_2378d}. In these figures, we also observed two parts of the jet: a front and a wake.

This simulation showed the following characteristics: a Central BH, a thin accretion disk and a relativistic with substructures front, wake and lobe. According to the unified theory of Active Galactic Nuclei (AGN) (reference), the lobes are in the fronts of jets produced by a supermassive BH at the center of the galaxy. The jets extend perpendicularly to the central accretion disk which surrounds the BH. Thus, our simulations produced elements compatible with AGN structures. This shows the potential of the developed code for the study of AGN, however, for such statement are needed simulations with real data.

So, with the second scenario adjustments, we have been able to simulate the jets formation, from the process of accretion disk, passing by the ejection and reaching the substructure lobe.

\section{Conclusions}

In this work, we were able to simulate the formation of relativistic collimated jets and lobes from the accretion disk rotating around a BH.

For this, we use a formalism for the GRMHD equations with axial symmetry in a fixed Schwarzschild metric.

We got stable numerical solutions with good behaviour towards the analytical solution for decreasing spatial domain cells, using central scheme free of Riemann Solvers, namely Two-dimensional Nessyahu-Tadmor method.

The first simulation, we started with the accretion disk 10,000 times denser than the magnetosphere. We observed the disk in falling into the BH, the matter deposited around the BH and finally the ejection of a jet up to a distance of $16r_{S}$. In this case, it has used a $630\times 630$ mesh which allow one see the formation of substructures along the jet which reaches relativistic speeds.

Comparisons were performed between numerical solutions at different meshes $210\times 210$, $630\times 630$ and $840\times 840$. Thus, it was verified stable solutions and with no significant differences between the solutions obtained in the meshes $630\times 630$ and $840\times 840$.

In the firs scenario, now with $210\times 210$ grid, we performed the jet until it reaches our outer radius, that is, until jet gets to $20r_{S}$. In this case, we observed the formation of a substructure at distance of almost $20r_{S}$.

So, in the second scenario, a new adjustment was realised. We set the initial accretion disk 100,000 times denser than the magnetosphere and, with this, we have gotten simulations with formation of substructure named Lobe, close to the disk not on the axis.

\section*{Acknowledgments}
We thank the anonymous referee for important indications of errors and mistakes. ROG thanks N. Kameswara Rao for some helpful suggestions and H. C. Bhatt for a critical reading of the original version of the paper. SRO acknowledges tips and references made by Luis Lehner at Perimeter Institute. 





\bsp

\label{lastpage}

\end{document}